\documentclass[12pt,fleqn]{article}

\usepackage{amsmath,amsfonts,bm,amssymb}
\usepackage{amsthm}
\usepackage{amsthm}

\usepackage{eucal}
\usepackage{eufrak}

\usepackage[a4paper]{geometry}

\usepackage{url}

\usepackage{multirow}
\usepackage{rotating}

\usepackage{graphicx}
\graphicspath{{figures/}} 

\newcommand{\remark}[1]{\marginpar{\scriptsize#1}}
\renewcommand{\remark}[1]{\marginpar{}}   

\long\def\symbolfootnote[#1]#2{\begingroup\def\thefootnote{\fnsymbol{footnote}}\footnote[#1]{#2}\endgroup}

\newcommand{\captionfonts}{\footnotesize}

\makeatletter  
\long\def\@makecaption#1#2{%
  \vskip\abovecaptionskip
  \sbox\@tempboxa{{\captionfonts #1: #2}}%
  \ifdim \wd\@tempboxa >\hsize
    {\captionfonts #1: #2\par}
  \else
    \hbox to\hsize{\hfil\box\@tempboxa\hfil}%
  \fi
  \vskip\belowcaptionskip}
\makeatother   

\def\ev #1{\left\langle #1 \right\rangle}

\def\MO{\mathrm{MO}}
\def\CA{\mathrm{CA}}
\def\LO{\mathrm{LO}}
\def\MOO{\mathrm{MO}^0}
\def\CAO{\mathrm{CA}^0}
\def\LOO{\mathrm{LO}^0}

\begin{document}
 
\title{\LARGE How does the market react to your order flow?}

\author{B. T\'oth\,$^{\textrm{a}, \textrm{b}}$, Z. Eisler\,$^{\textrm{a}}$, F. Lillo\,$^{\textrm{b},\textrm{c},\textrm{d}}$, \\
J. Kockelkoren\,$^{\textrm{a}}$, J.-P. Bouchaud\,$^{\textrm{a}}$, J.D. Farmer\,$^{\textrm{b}}$}

\date{\today}
\maketitle
\small
\begin{center}
  $^\textrm{a}$~\emph{Capital Fund Management, 6, blvd Haussmann 75009 Paris, France}\\
  $^\textrm{b}$~\emph{Santa Fe Institute, 1399 Hyde Park Rd., Santa Fe NM 87501}\\
  $^\textrm{c}$~\emph{Scuola Normale Superiore di Pisa, Piazza dei Cavalieri 7, I-56126 Pisa, Italy}\\
  $^\textrm{d}$~\emph{Dipartimento di Fisica, Universit\`a di Palermo, Viale delle Scienze, I-90128 Palermo, Italy}\\
\end{center}
\normalsize

\vspace{0.3cm}

\begin{abstract}
We present an empirical study of the intertwined behaviour of members in a financial market. 
Exploiting a database where the broker that initiates an order book event can be identified, we decompose 
the correlation and response functions into contributions coming from different market participants and study how their behaviour 
is interconnected. We find evidence that (1) brokers are very 
heterogeneous in liquidity provision -- some appear to be primarily liquidity providers while others are primarily liquidity takers.  
(2) The behaviour of brokers is strongly conditioned on the actions of {\it other} brokers.  In contrast brokers are only weakly influenced by the impact of their own previous orders. (3) The total impact of market orders is the result of a subtle compensation between 
the same broker pushing the price in one direction and the liquidity provision of other brokers pushing it in the opposite direction.  
These results enforce the picture of market dynamics being the result of the competition between heterogeneous participants, interacting 
to form a complex market ecology.
 \end{abstract}

\smallskip

\smallskip


\section{Introduction}
\label{intro}

Empirical studies of order flow and market impact have recently boomed due to the availability of high frequency data, where all market events (trades, limit orders, cancellations) are recorded. 
These data sets allow one to investigate many interesting statistical regularities at the order book level, and shed light on the price formation mechanisms. One of the most interesting results 
established in the recent literature concerns the long-ranged correlated nature of the order flow, and a detailed understanding of the impact of individual transactions on prices (see \cite{bouchaud2009} for a recent review, and references therein). 

However, most empirical studies to date rely on a purely anonymous order flow: trades, limit orders and cancellations cannot be traced to a particular agent in the market. In order to 
access this data, some special agreement must be reached with exchanges, which regulators allow or even promote in certain conditions 
(for example in order to investigate the role of  ``high-frequency  traders'' in the market, see the interesting recent papers \cite{kirilenko2010,menkveld2011}). 

The data that we exploit here is unfortunately not as detailed, but allows us to identify the activity of market members of the LSE 
(London Stock Exchange) in the period May, 2000 -- December, 2002. Since members are often brokers who act on the behalf of many final clients, the granularity of the order flow is quite coarse, still some interesting conclusions can be drawn from this data, as has been shown in \cite{lillo2008,moro2009,toth2010}. Note that since most members also act as brokers, we will throughout use the 
word ``broker" as synonymous with ``member" and we will use these terms  interchangeably.  
Here, we want to adapt a formalism introduced by some of us in Ref. \cite{Eisler2010} to investigate the 
correlation and impact of various types of order book events. In that paper, events were broken down into six categories: market orders, limit orders and cancellations, and for each type whether the event 
immediately changes the midpoint price or not. In principle, further categories can be envisaged, and here we add the brokerage code as an additional tag. 

Using this decomposition, our goal is to elicit how the {\it interaction} between {\it heterogeneous} actors in the market results in a {\it subtle ecology} \cite{handa1998,bouchaud2004}, with  measurable effects in the price dynamics. We first use the 
type of order (limit orders vs. market orders) as conditioning variable to check for heterogeneity among the brokers. We find that it is possible to distinguish between ``liquidity providers'' 
(placing predominantly limit orders) and ``liquidity takers'' (placing predominantly market orders) in the market. We furthermore find that there is a clear difference between the impact of price-changing 
limit orders and price-changing market orders, depending on the category of brokers. Next we look at the temporal dynamics of the different types of orders. Further conditioning on whether a limit order 
or a market order changed the mid-price or not, we find that the actions of brokers are strongly conditioned on the actions of other brokers. Brokers react to the same price change in a different way 
whether the price change is a result of their own actions, or results from the action of other brokers. Finally we decompose the total impact of a given type of order book event into a contribution from 
the very broker that caused the initial event and a contribution from all other brokers. We find that these two contributions {\it very nearly offset each other}, leading to a total impact that is nearly 
constant in time and {\it much smaller than either of these contributions}. This is the central result of this paper, which confirms the dynamical liquidity picture put forth in 
\cite{lillo2004,bouchaud2006,farmer2007,gerig2007,bouchaud2009,Eisler2010}, according to which the highly persistent sign of market orders 
must be buffered by a fine-tuned counteracting limit order flow in order to maintain statistical 
efficiency (i.e. that the price changes are close to unpredictable, in spite of the long-ranged correlation of the order flow, see also \cite{sqrt}).We believe that the quantitative result presented here is a very important ingredient to understand the dynamics of markets, since it explicitly demonstrates that the stability of markets relies on a rather precise balance between liquidity taking and liquidity providing, 
and that small fluctuations of one or the other can lead to micro-liquidity crises and price jumps \cite{joulin2008,bouchaud2011}.

\section{Data and notation}
\label{notations}

In this paper we analyze data on 7 of the most liquid stocks traded at the LSE during the period May, 2000 -- December, 2002\footnote{The studied stocks are AstraZeneca (AZN), BHP Billiton (BLT), 
Lloyds Banking Group (LLOY), Prudential (PRU), Rentokil Initial (RTO), Tesco (TSCO), Vodafone Group (VOD).}. Our data set contains all on-book events for the stocks. We only consider the usual trading time 
between 8:00--16:30, all other periods are discarded. The unique feature of our data set is that each order is characterized by a membership
code identifying the initiator of the order. These 
codes uniquely represent the member firms of the LSE even if we are not able to identify the firms by name. (Actually the brokerage codes in our data are reshuffled at the beginning of each 
month. However, since most of the results we show happen at the intraday scale, we can ignore the effect of the reshuffling.)
The activity level of brokerages is very heterogeneous. For example, in a typical month for AZN the 5 most active market members are responsible for 40-50\% of transactions and the 15 most active ones are 
responsible for 80-90\% of transactions. Thus the trading activity is strongly concentrated in a relatively small number of member firms.
In Table \ref{table1} we show summary statistics of the stocks and the brokerages. We present the number of events for each stock, the ratio of the tick size to the average price, the typical (average) number of 
active brokers.  In a given month we define $\alpha_i$ as the fraction of 
trades initiated by broker $i$.   The typical Gini coefficient of $\alpha_i$ is about $0.8$, and the typical standard deviation of $\log_{10}\alpha_i$ is a little more than one, indicating that the typical 
difference between the activity of two brokers chosen at random is more than an order of magnitude.

\begin{table}
\begin{center}
\begin{tabular}{| c | r| c | c | c | c |}
\hline
ticker &  number  & tick/avg price &  typical number & Gini  & $\mathrm{std}(\log_{10}\alpha_i)$\\
 & of events & [$\times10^{-4}$] & of brokers & coeff. & \\
\hline
AZN & 1,846,922 & 3.49 & 95 & 0.80 & 1.12\\ 
BLT & 958,573 & 7.69 & 69 & 0.75 & 1.03\\ 
LLOY & 1,761,548 & 7.84 & 104 & 0.79 & 1.13\\ 
PRU & 1,312,220 & 7.44 & 88 & 0.77 & 1.10\\ 
RTO & 714,665 & 11.07 & 65 & 0.75 & 1.02\\ 
TSCO & 1,175,353 & 10.68 & 87 & 0.78 & 1.10\\ 
VOD & 2,712,084 & 15.23 & 134 & 0.82 & 1.13\\ 
\hline
\end{tabular}
\end{center}
\caption{\label{table1}Summary statistics for the data we study here.  The first three columns are the ticker, the total number of events and the ratio of the tick size to the average price. Letting 
$\alpha_i$ be the fraction of the trades during a given month initiated by broker $i$, the other fourth column shows the typical (average) number of active brokerages (defined to be those with 
$\alpha_i > 0.01$). The fourth column is the Gini coefficient of $\alpha_i$, and the sixth is the standard deviation of $\log_{10} \alpha_i$.}
\end{table}

Following Ref. \cite{Eisler2010} we will analyse time series of order book events.
We use the name ``event'' for any change in the order book that modifies the bid or ask price or the volume quoted at these prices. Events will be used as the unit of time. Since there can be many 
events between two transactions this notion of  ``event time'' is similar 
but more fine-grained than the notion of transaction time used in many papers. The price just before the $t^{th}$ event is defined as the midpoint price $p_t$,  i.e. the average of the best ask and 
best bid quote. We use ticks as the units of price. The type of event at time $t$ will be denoted by $\pi_t$.  The upper index $'$ (``prime'') denotes that an event changed the price $p_t$, and the 
upper index 0 that it did not. The possible events are:\begin{itemize}
\item 
$\MOO$ is an effective market order\footnote{
By {\it effective market order} we mean any event that generates immediate transactions with existing orders in the limit order book.}
that does not change the price.
\item
$\MO'$ is an effective market order that does change the price.
\item
$\LOO$ is a limit order at the current bid or ask (which does not change the price).
\item
$\LO'$ is a limit order inside the spread (which does change the price).
\item
$\CAO$ is cancellation at the bid or ask that does not remove all the volume quoted (and thus does not change the price).
\item
$\CA'$ is a cancellation at the bid or the ask that does remove all the volume quoted (and thus does change the price).
\end{itemize}
Abbreviations without the upper index ($\MO$, $\CA$, $\LO$) refer to events whether or not they change the price.  
We will not explicitly consider limit orders and cancellations inside the order book, because they do not have an immediate effect on the best quotes. The event at time $t$ is given a sign $\epsilon_t$ 
according to its expected long-term effect on the price. For market orders and limit orders this corresponds to order signs, i.e., $\epsilon_t = 1$ for buy orders and $-1$ for sell orders. Cancellation 
of a sell limit order has $\epsilon_t = 1$ while cancellation of a buy limit order has $\epsilon_t = -1$ (the sign is reversed because the effect on the price is in the opposite direction).

We will use the indicator function $I(\mathrm{A})$ which is defined as  $I(\mathrm{A})=1$ if the condition $\mathrm{A}$ is true and  $I(\mathrm{A})=0$ otherwise. For example, the indicator variable 
$I(\pi_t=\pi)$ is $1$ if the event at time $t$ is of type $\pi$ and zero otherwise. The unconditional probability of the event type $\pi$ is by definition $P(\pi)=\ev{I(\pi_t=\pi)}$. The market member 
acting at time $t$ will be denoted by $b_t$, and the indicator function $I(b_t=b)$ is $1$ if the broker acting at time $t$ is $b$ and zero otherwise.

As discussed in Ref. \cite{Eisler2010}, the use of the indicator functions simplifies the formal calculation of conditional expectations. For example, if a quantity $X_{\pi, t}$ depends on the event type $\pi$ and and the time $t$, then its conditional expectation at times of $\pi$-type events is
\begin{eqnarray}
\ev{X_{\pi_t, t}~|~\pi_t=\pi}=\frac{\ev{X_{\pi,t}I(\pi_t=\pi)}}{P(\pi)}.
\end{eqnarray}

The average behaviour of the price $\ell$ time steps after an event of a particular type $\pi_1$ defines the corresponding response function (or average impact function) \cite{bouchaud2004}

\begin{eqnarray}
{\cal R}_{\pi_1}(\ell)=\frac{\ev{(p_{t+\ell}-p_{t})I(\pi_t=\pi_1)\epsilon_t}}{P(\pi_1)}.
\label{eq_resp_def}
\end{eqnarray}

\noindent This response can be divided into a part that is the response due to the actions of the same broker as the one active at time $t$ (${\cal R}^{\mathrm{same}}_{\pi_1}(\ell)$), and the response 
due to other brokers (${\cal R}^{\mathrm{diff}}_{\pi_1}(\ell)$).

\noindent The response due to the same broker can be written

\begin{eqnarray}
{\cal R}^{\mathrm{same}}_{\pi_1}(\ell)=\frac{\ev{\sum_{t'=t}^{t+\ell-1}(p_{t'+1}-p_{t'})I(b_{t'}=b_{t})I(\pi_t=\pi_1)\epsilon_t}}{P(\pi_1)}.
\label{eq_resp_same_def}
\end{eqnarray}

\noindent ${\cal R}^{\mathrm{same}}_{\pi_1}(\ell)$ is the expected price change between time $t$ and $t+\ell$ caused by the further 
actions of the same broker that acted at time $t$ and ignoring all other brokers (since $I(b_{t'}=b_{t})$ picks out only events from the same broker).
Conversely the response that is only due to other brokers can be written

\begin{eqnarray}
{\cal R}^{\mathrm{diff}}_{\pi_1}(\ell)=\frac{\ev{\sum_{t'=t}^{t+\ell-1}(p_{t'+1}-p_{t'})I(b_{t'}\neq b_{t})I(\pi_t=\pi_1)\epsilon_t}}{P(\pi_1)}.
\label{eq_resp_other_def}
\end{eqnarray}

\noindent Trivially ${\cal R}^{\mathrm{same}}_{\pi_1}(\ell)+{\cal R}^{\mathrm{diff}}_{\pi_1}(\ell)={\cal R}_{\pi_1}(\ell)$.
Furthermore, we can define the different contributions to the response ${\cal R}^{\mathrm{same}}_{\pi_1}(\ell)$ coming from the possible $\pi_2$ types of events occurring at $t'$ as
\begin{eqnarray}
{\cal R}^{\mathrm{same}}_{\pi_1, \pi_2}(\ell)=\frac{\ev{\sum_{t'=t}^{t+\ell-1}(p_{t'+1}-p_{t'})I(\pi_{t'}=\pi_2) I(b_{t'}=b_{t})I(\pi_t=\pi_1)\epsilon_t}}{P(\pi_1)},
\label{eq_resp_same_contributions_def}
\end{eqnarray}
and similarly for ${\cal R}^{\mathrm{diff}}_{\pi_1, \pi_2}(\ell)$.

\noindent In a similar way one can define the correlation function of order signs\footnote{Note that the correlation function that we use is the conditional expectation $\ev{\epsilon_t\epsilon_{t+\ell} ~|~\pi_t=\pi_1,~\pi_{t+\ell}=\pi_2}$ and is not normalized between $[-1,1]$.}

\begin{eqnarray}
C_{\pi_1,\pi_2}(\ell)=\frac{\ev{I(\pi_{t+\ell}=\pi_2)\epsilon_{t+\ell} I(\pi_t=\pi_1)\epsilon_t}}{P(\pi_1)P(\pi_2)},
\label{eq_corr_def}
\end{eqnarray}

\noindent which again can be divided as the sum of the correlation between events initiated by the same broker ($C^{\mathrm{same}}_{\pi_1,\pi_2}(\ell)$) and the correlation between events initiated by 
different brokers ($C^{\mathrm{diff}}_{\pi_1,\pi_2}(\ell)$),

\begin{eqnarray}
C^{\mathrm{same}}_{\pi_1,\pi_2}(\ell)=\frac{\ev{I(\pi_{t+\ell}=\pi_2)\epsilon_{t+\ell} I(\pi_t=\pi_1)\epsilon_t I(b_{t+\ell}=b_{t})}}{P(\pi_1)P(\pi_2)},
\label{eq_corr_same_def}
\end{eqnarray}

\begin{eqnarray}
C^{\mathrm{diff}}_{\pi_1,\pi_2}(\ell)=\frac{\ev{I(\pi_{t+\ell}=\pi_2)\epsilon_{t+\ell} I(\pi_t=\pi_1)\epsilon_t I(b_{t+\ell}\neq b_{t})}}{P(\pi_1)P(\pi_2)}.
\label{eq_corr_other_def}
\end{eqnarray}

For simplicity, when talking about the response functions and the correlations, we will simply refer to ${\cal R}^{\mathrm{same}}_{\pi_1}(\ell)$ and $C^{\mathrm{same}}_{\pi_1,\pi_2}(\ell)$ as the 
contribution of the {\it same broker}, while we will refer to ${\cal R}^{\mathrm{diff}}_{\pi_1}(\ell)$ and $C^{\mathrm{diff}}_{\pi_1,\pi_2}(\ell)$ as the contribution of {\it other brokers}.

When showing our results, we averaged over all the 7 stocks studied. However, we checked all the results for each stock individually and found
that the results are very similar.

\section{Heterogeneity of broker liquidity provision}
\label{heterogeneity}

Are brokers homogeneous in the sense that an event from a given broker has the same statistical properties as an event from any other broker?  Or are brokers heterogeneous in the sense that their events 
have different statistical properties?  In this section we show that when it comes to liquidity provision brokers are very heterogeneous.

As is well known, the definition of liquidity is not unique, but in the present paper by liquidity we mean the volume in the order book. According to this definition limit orders provide liquidity and market orders take away liquidity from the market. Within the classic market microstructure models \cite{amihud1980, handa1996, Madhavan, Hasbrouck, menkveld2011}, such as the Glosten-Milgrom model \cite{glosten1985}, investors are classified into two categories: informed traders and market makers (often a third category, noise traders are added).
Informed investors are supposed to
possess superior information and use market orders to exploit their information immediately. Some other participants specialize in market making activities, who provide liquidity to both the buy and 
sell side and attempt to make a profit from the bid-ask spread. In this view, informed traders are liquidity takers, while liquidity providers are market makers. However, in modern automated markets (such as the LSE) this distinction is not obvious since any investor can use market or limit orders to trade and therefore act alternatively as a liquidity taker or as a provider. 
Recent empirical research, such as the one in Ref. \cite{kirilenko2010}, has indeed found that different types of investors, such as fundamental investors and high frequency traders, use a similar mixture of limit and market orders. 

\begin{figure}[h]
\begin{center}
\includegraphics[width=0.49\textwidth, angle=0]{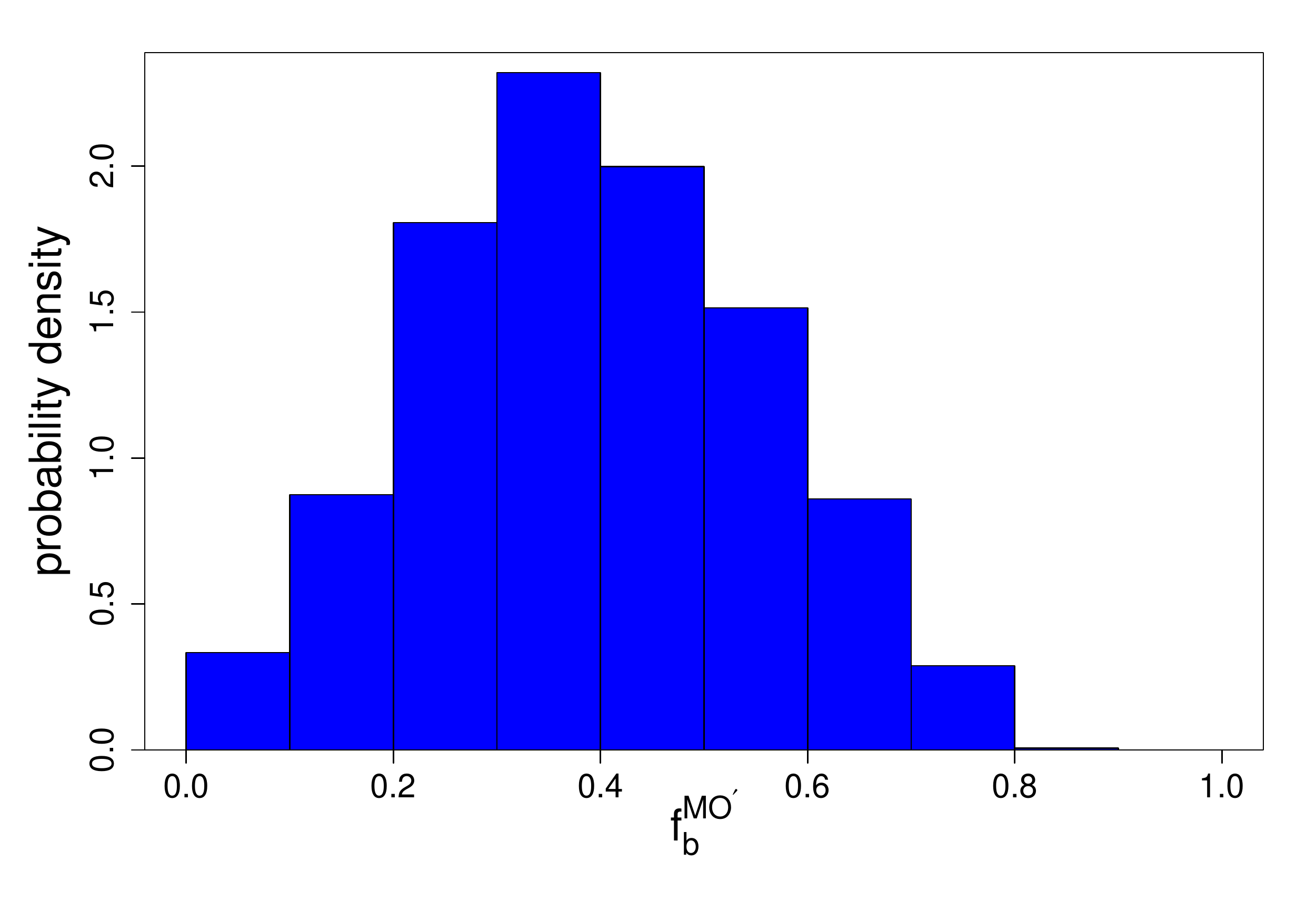}
\includegraphics[width=0.49\textwidth, angle=0]{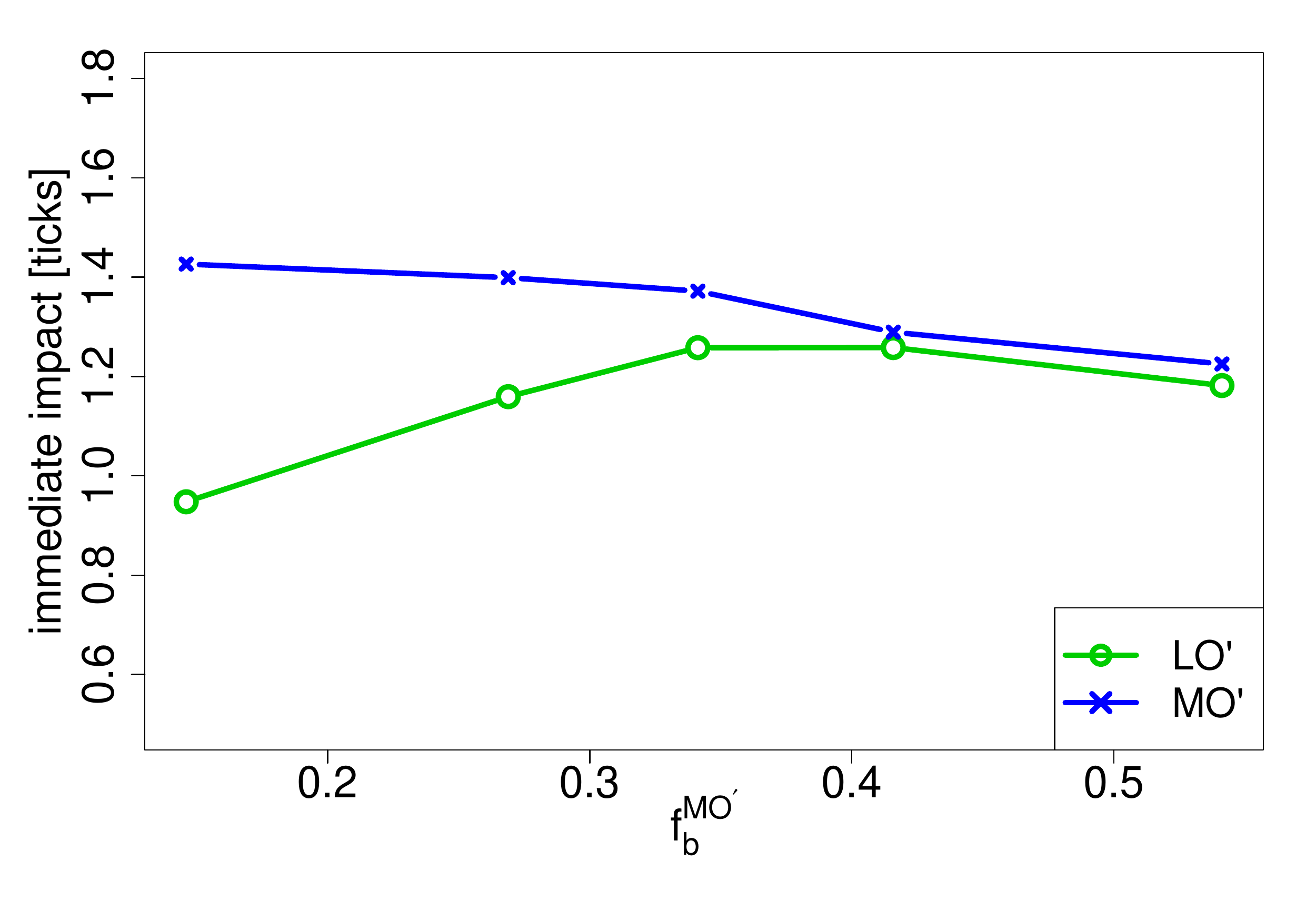}
\caption{(left) The distribution of the ratio $f^{\MO'}_{b}$, defined in Eq.~(\ref{feq}), which takes on low values for brokers who tend to be liquidity providers and higher values for brokers that tend to be 
liquidity takers.  We only include brokers whose number of trades is at least $1\%$ of the total in a given month. (right) The immediate impact ${\cal R}_{\pi}(\ell=1)$ for price changing limit orders $\pi=\LO'$ 
and $\pi=\MO'$ orders is plotted as a function of $f^{\MO'}_{b}$, the fraction of price changing market orders.   To reduce statistical fluctuations the data are binned into five groups according to $f^{\MO'}_{b}$.}
\label{fig_brokers}
\end{center}
\end{figure}

Here we want to investigate whether individual brokers show a heterogeneous profile of liquidity provision. To this end we measure for each broker the number of price changing market orders over the 
total number of price changing orders.  Note that we consider only price changing  orders (and not all MO and LO), since these are orders with an immediate effect on the price. This we do in order to 
include less bias in the statistics due to orders being placed and cancelled almost immediately. More formally, let $\#MO'_b$ be the number of price changing market orders placed by broker $b$ in a 
given month, and similarly $\#LO'_b$ be the number of price changing limit orders placed by broker $b$ in that month.  The fraction
\begin{equation}
f^{\MO'}_{b} = \frac{\#\MO'_b}{\#\MO'_b+\#\LO'_b}
\label{feq}
\end{equation}
is the fraction of price changing market orders placed by broker $b$ in a given month (we use a month because the brokerage codes are shuffled every month).  A high value of $f_b^{\MO'}$ thus implies that 
a broker tends to be a liquidity taker and a low value of $f_b^{\MO'}$ implies that she tends to be a liquidity provider.
The distribution\footnote{
Note that the distribution is not weighted by the size of the broker.}
of the average values of $f^{\MO'}_{b}$ is shown in the left panel of Fig.~(\ref{fig_brokers}). This distribution is extremely broad, with average values of $f_b^{\MO'}$ ranging from $f_b^{\MO'} \approx 0$, 
indicating a broker that acts as a liquidity provider, using limit orders almost exclusively, to $f_b^{\MO'} \approx 0.8$, indicating a broker that acts
as a liquidity taker, predominantly using market orders.

Statistical testing shows that the wide variation in $f_b^{\MO'}$ is due to real heterogeneity among the brokers rather than statistical fluctuations.  If we define a series with values of one or zero 
depending on whether a price changing order is an effective market order or a limit order, the characteristic time for the autocorrelation to decay into the noise level is less than ten price changing 
orders.  This enables us to estimate the standard deviation of the $f^{\MO'}_{b}$ value for each broker by drawing block bootstrap samples for all brokers in all periods (with block sizes of 10 orders). 
The typical standard deviation for the $f^{\MO'}_{b}$ values is less than $0.04$.  Thus the broad distribution of values of $f_b^{\MO'}$ shown in Fig.~(\ref{fig_brokers}) is almost entirely real variation, 
corresponding to heterogeneous behaviour across brokers. We find that $f^{\MO'}_{b}$ does not show significant dependence on the size of the broker (defined by his total number of orders of any type), so there is no 
simple relation between transaction volumes and the use of market vs. limit orders that would reflect, for example, some systematic difference of information between large and small brokers.

Nonetheless, the results summarized in the right panel of Fig.~(\ref{fig_brokers}) suggest that brokers with different levels of liquidity provision specialize in different types of execution.  We show 
the average immediate impact ${\cal R}_{\pi}(\ell=1)$ of an order of type $\pi=\LO'$ and $\pi=\MO'$ as a function of $f^{\MO'}_{b}$.  To reduce the statistical variation we bin the brokers into five groups 
according to $f^{\MO'}_{b}$, and plot the average value of ${\cal R}_{\pi}(\ell=1)$ in each bin against the average value of $f^{\MO'}_{b}$ for that bin.  The impact for market orders decreases slightly 
with $f^{\MO'}_{b}$, indicating that brokers who use market orders more frequently get better execution.  For limit orders, in contrast, the lowest impact is for low values of $f^{\MO'}_{b}$, indicating 
that brokers who use limit orders more frequently get better execution.  For the top two quintiles of $f^{\MO'}_{b}$ there is essentially no difference between the impact of limit orders and market orders, 
while for the lowest quintile the difference is dramatic -- limit orders have an impact of about a tick, while market orders have more than 1.4 ticks of impact. This difference is statistically robust. 
Details of the statistical tests can be found in Appendix \ref{app}.

These observations are compatible with the standard classification made in the microstructure literature and reviewed above. In particular, brokers who are predominantly providing liquidity tend to transact passively with less information so that their limit orders have less impact than those of directional traders. Still, in the rare cases where these liquidity providers use market orders, e.g. to flatten their positions at the end of the day, they tend to execute under unfavourable circumstances, e.g. under more time pressure to execute a large quantity, and therefore have a larger impact.

This interpretation suggests that the correlation of the order flow should be quite different for brokers with low values of 
$f^{\MO'}_{b}$ and for brokers with large values of $f^{\MO'}_{b}$. If the former category really acted primarily as market makers, their order flow
should be significantly {\it negatively} correlated in time, whereas liquidity takers would contribute to the overall positive correlation of the order flow reported below. 
However, we did not find any significant difference between the groups. Because of this we cannot conclude that brokers predominantly providing liquidity act as traditional market makers. 
This might be partly due to the fact that even directional traders make extensive use of limit orders (as it was pointed out in, e.g. \cite{bouchaud2009}). 

Our only firm conclusion is therefore that there is a significant heterogeneity among brokers in the way they are using market orders versus limit orders when 
executing a large metaorder. The right panel of Fig.~(\ref{fig_brokers}) suggest that brokers who use a given type of order more often are more skillful 
at using this type of order, in the sense that they impact less the price by doing so. (We assume here that a smaller impact means a better execution price, which might not always be the case).

\section{Regularities in order placement}

The above analysis can also be used to gain insight into the origin of long-ranged correlation in the sign of orders. Using the above formalism, we can break up this correlation into different contributions, depending 
on whether or not the event is price changing and whether the broker is the same or different.  As we will see, the response of brokers to their own price changes is quite different than their response to the price 
changes of others.

\begin{figure}[h]
\begin{center}
\includegraphics[width=0.49\textwidth, angle=0]{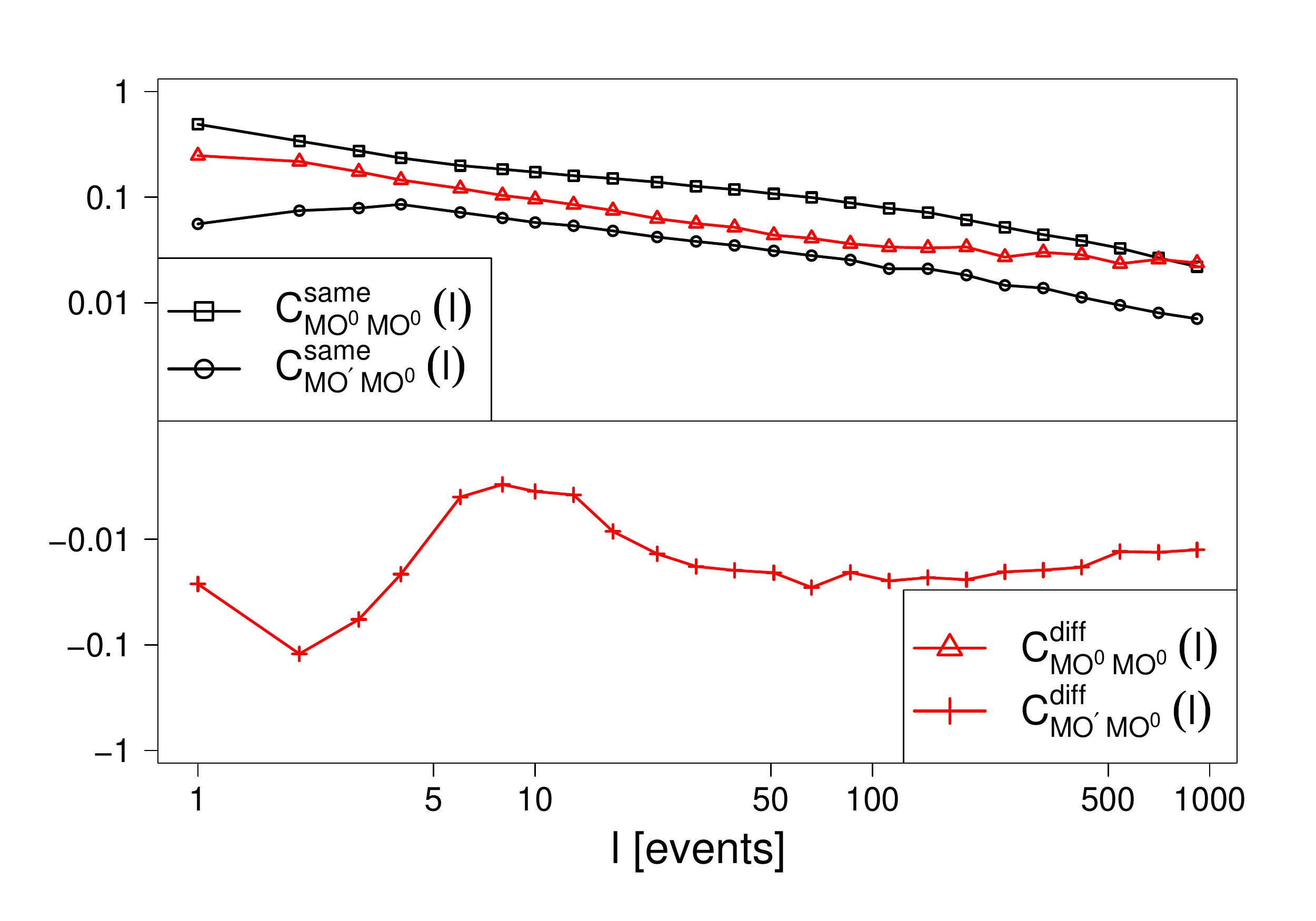}
\includegraphics[width=0.49\textwidth, angle=0]{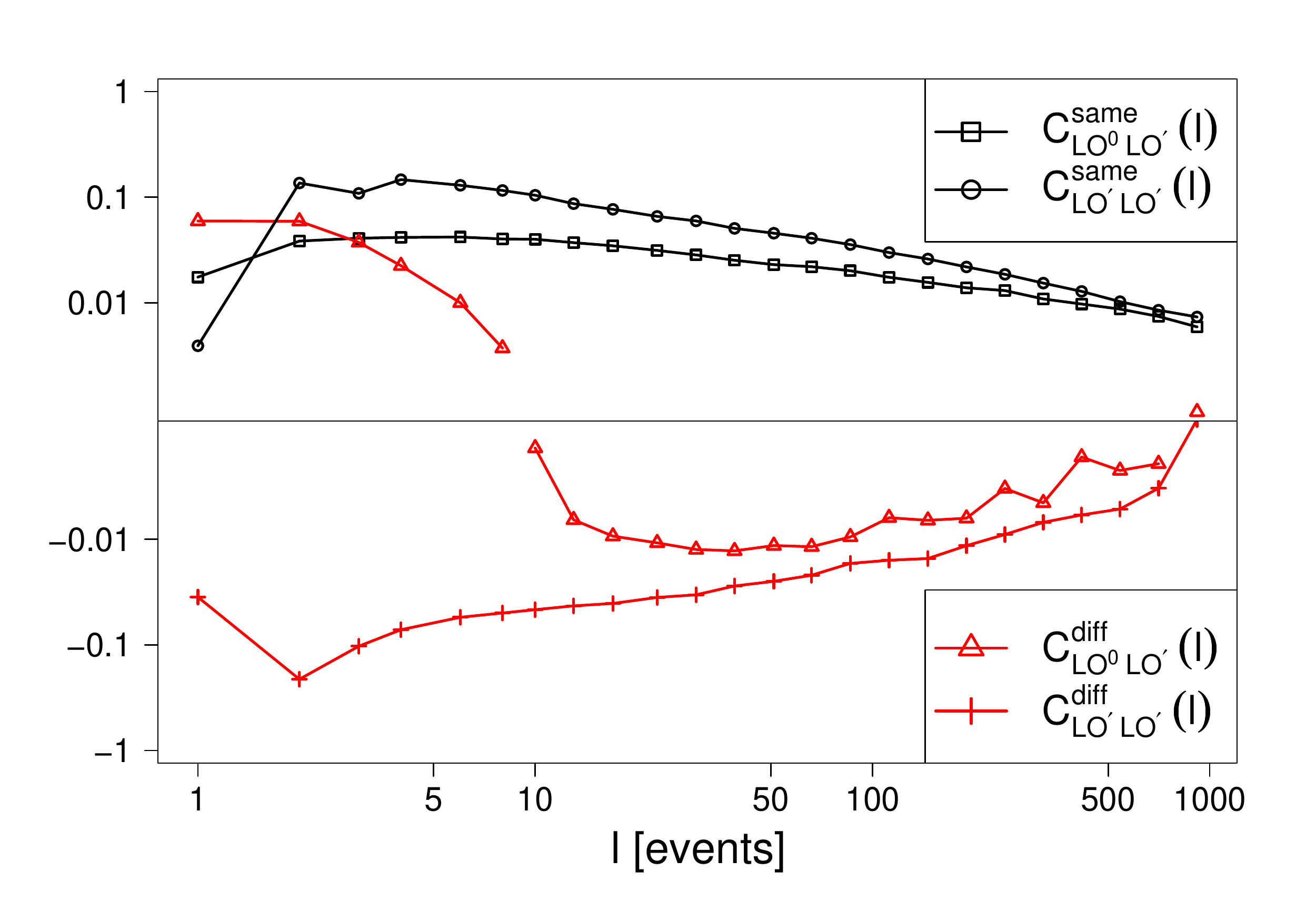}
\caption{Correlation functions between events. (left) The correlation $C_{\MOO,\MOO}(\ell)$ between non-price-changing market orders at two different times, and  the correlation $C_{\MO',\MOO}(\ell)$ between a 
price-changing market order and a non-price-changing market order.  In each case we present results conditioned on the same broker for the two events vs. a different broker for the two events. (right) The 
correlation $C_{\LOO,\LO'}(\ell)$ for a non-price changing and a price changing limit order, and the correlation $C_{\LO',\LO'}(\ell)$ for two price changing limit orders.}
\label{fig_corr}
\end{center}
\end{figure}

In the left panel of Fig.~(\ref{fig_corr}) we plot some relevant correlations for market orders, in particular $C_{\MOO,\MOO}(\ell)$ and $C_{\MO',\MOO}(\ell)$, conditioning on the same broker and on different brokers. 
It can be seen that:
\begin{itemize}
\item i) The autocorrelations $C^{\mathrm{same}}_{\MOO,\MOO}(\ell)$, $C^{\mathrm{same}}_{\MO',\MOO}(\ell)$, and $C^{\mathrm{diff}}_{\MOO,\MOO}(\ell)$ all behave similarly.  They are positive for lags up to more than 
500, and they all decay roughly as power laws, $\ell^{-\gamma}$ with $\gamma \approx 0.5$. This is very close to the exponent found 
in \cite{lillo2004,bouchaud2004} for the unconditional autocorrelation of market order sign.  The three autocorrelations differ in amplitude:  $C^{\mathrm{same}}_{\MOO,\MOO}(\ell)$ is largest, with a value of roughly 
$0.8$ for lag one, then $C^{\mathrm{diff}}_{\MOO,\MOO}(\ell)$, with a value of roughly $0.2$ for lag one, and finally $C^{\mathrm{same}}_{\MO',\MOO}(\ell)$, with a value less than $0.1$ at lag one.  They all decay 
roughly in the same way across the entire range (see also \cite{Toth2011}).
\item ii)  In contrast, the autocorrelation $C^{\mathrm{diff}}_{\MO',\MOO}(\ell)$, which measures the response of another broker to a price changing market order, is weakly but consistently negative and shows no 
clear pattern of decay.  Furthermore its behaviour is completely different to that of $C^{\mathrm{diff}}_{\MOO,\MOO}(\ell)$. \end{itemize}

Stated in different terms, the first set of results imply that the response to a non-price-changing market order is always the same:  Whether subsequent orders come from the same broker or a different broker, 
subsequent order placements tend to be of the same sign.  This is in contrast to the response to a price-changing market order.  In this case the same broker tends to keep placing orders of the same sign as her 
original order, while other brokers tend to place orders of opposite signs to the original order. 

To asses the statistical significance of the correlations we constructed the following test. We randomly reshuffled the time series of order signs, $\epsilon_t$, and measured the correlation of the reshuffled series. Since by reshuffling we purposefully destroy all correlations, we can expect that the correlation level of the reshuffled time series is the noise level of our correlation measure. Carrying out $1000$ such reshuffling experiments and measuring the correlation function in each case, for the absolute value of the autocorrelation we get an average of $\tilde{C}\approx 5\cdot 10^{-4}$, that is $2-3$ orders of magnitude lower than the measured correlation values. The 99\% quantile
of the absolute value of the autocorrelations is $\approx 2\cdot 10^{-3}$.
This tells us that all the correlation curves in Fig.~(\ref{fig_corr}) are indeed significant.

The observation of long-memory autocorrelation functions for orders placed by the same broker provides additional evidence that at the brokerage level the long-memory of order flow is primarily driven by the splitting of large metaorders into small pieces, as posited in \cite{lillo2005,bouchaud2009} (see also \cite{gerig2007,Toth2011} for additional empirical evidence supporting this hypothesis). The theoretical motivations for order splitting were first 
discussed by Kyle \cite{kyle1985} (note however that Kyle's model does {\it not} generate correlation in the sign of the trades!). 
Ref. \cite{obizhaeva2005} discusses an optimal execution strategy of block trades, and conclude that order splitting is indeed optimal. 
Recently some of us \cite{sqrt} proposed a dynamical theory of market liquidity that predicts a vanishingly small  volume around the current price resulting in a need to split large orders.

The autocorrelation of orders placed by the same broker decays slowly whether or not the original 
market order causes a price change.  Given that a price change in the same direction is unfavourable (a buyer does not want the price to rise), this is a bit surprising at first sight.  However, note that 
although there is long-memory in both cases, the magnitude of the correlation is roughly a factor of eight smaller for orders issued by the same 
broker.

The fact that order flow by the same broker continues in the same direction suggests that parent orders (or `metaorders') are 
executed to a large extent independently of the price change, even when this change is caused by trading. From a behavioural finance perspective, agents doing splitting appear to act like ``noise traders'', i.e. 
they adapt their own order flow very little to the effect of their own recent trading. This is the essence of the model proposed in \cite{lillo2005}. 
This, we believe, is either because agents neglect impact altogether (it is after all a small effect compared to the volatility of prices), or because they have already factored in the impact of their trades in their 
estimates of trading costs and thus are rationally following their plan to let their order run until executed.

The fact that the autocorrelation in response to a non-price-changing market order is also positive, even for different brokers, suggests herding behaviour.  This could happen in two ways:  a) one is that after a 
non-aggressive  market order $\MOO$, other brokers jump on the bandwagon, thinking there might be some information in the initial trade, or alternatively, b) the other brokers might be responding to the same information signals as the original broker, but with a slight lag.

Surprisingly, though, if the original market order is price-changing, the sign of this effect is reversed.  
In this case the original market order triggers the activity of ``other brokers'' with the opposite sign. 
The observation of a price rise converts the other brokers (or at least a majority of them) from buyers to sellers, or from sellers to buyers.  This is compatible with the idea of a large liquidity buffer that reveals itself as soon as the price changes (see \cite{bouchaud2006,sqrt}), which seems to be enough to overwhelm the herding effect that is seen when there is no price change.   

Similar behaviour is seen for limit orders as well when the broker is the same, but the situation is somewhat altered when the broker is different.  The right panel of Fig. \ref{fig_corr} shows $C_{\LOO,\LO'}(\ell)$ 
and $C_{\LO',\LO'}(\ell)$. Again, 
we find that $C^{\mathrm{same}}_{\LOO,\LO'}(\ell)$ is similar to $C^{\mathrm{same}}_{\LO',\LO'}(\ell)$. In this case, however, there is no herding on the part of other brokers since both  
$C^{\mathrm{diff}}_{\LOO,\LO'}(\ell)$, $C^{\mathrm{diff}}_{\LO',\LO'}(\ell)$ are negative (except at very short lags $\ell < 5$ for $C^{\mathrm{diff}}_{\LOO,\LO'}(\ell)$).  Thus at long time lags other brokers 
respond to limit orders by placing orders of the opposite sign, whether or not the original order was price-changing.

Even restricting to $\LO$'s and $\MO$'s and discarding $\CA$'s, one can define a total of 32 different correlation functions, whereas Fig. \ref{fig_corr} only shows 8 of these.  Each of these correlation functions answers a different question: conditioned to an event of type $\pi_1$ issued by a broker $b$, 
what is the excess probability that a broker $b'$ (with $b'=b$ or $b'\neq b$) issues an event of type $\pi_2$ with the same sign after $\ell$ 
trades? For lack of space, and because not all these correlations tell interesting stories, we choose to restrict here to three additional ones, shown in Fig. \ref{fig_corr_disaggregated}.\footnote{All
the correlation functions are available in the Appendix of the arXiv version of this paper, see http://arxiv.org/abs/1104.0587.}
The question they answer is the following: conditioned on a non-price-changing order by broker $b$, what is the excess probability that the {\it same broker} $b$ issues a price-changing order with the same sign after $\ell$ trades, when the original order is either a market order $\MOO$ or a limit order $\LOO$.  The four excess probabilities are given by  $P(\pi_2) C^{\mathrm{same}}_{\pi_1,\pi_2}(\ell)$ with $\pi_1=\MOO$ or $\LOO$ and $\pi_2=\MO'$ or $\LO'$, as shown in Fig. \ref{fig_corr_disaggregated}. (Note that $C^{\mathrm{same}}_{\LO',\LO'}(\ell)$  already appears in Fig. \ref{fig_corr}, right, so that we indeed only add three new quantities to the above.)

\begin{figure}[h]
\begin{center}
\includegraphics[width=0.49\textwidth, angle=0]{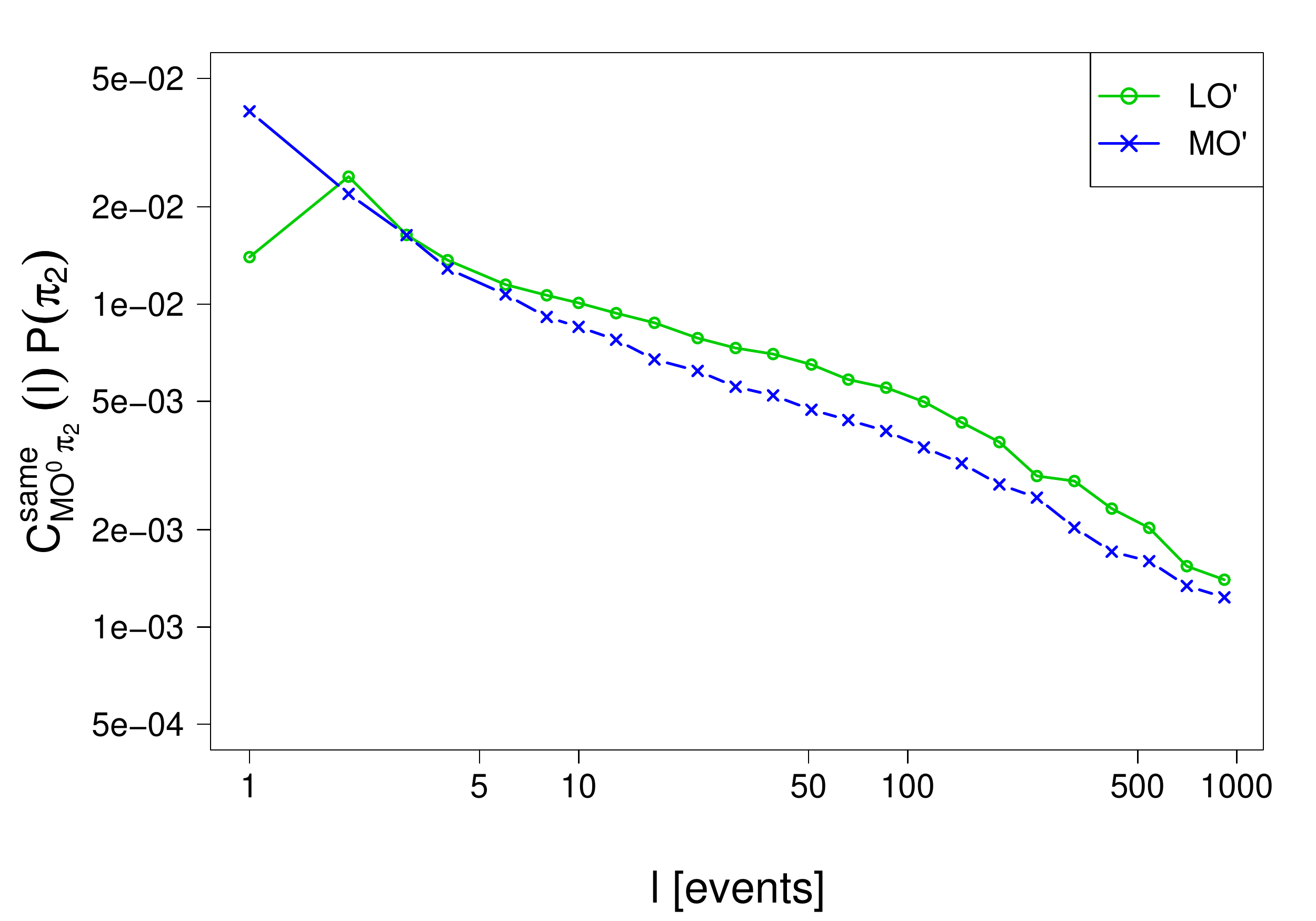}
\includegraphics[width=0.49\textwidth, angle=0]{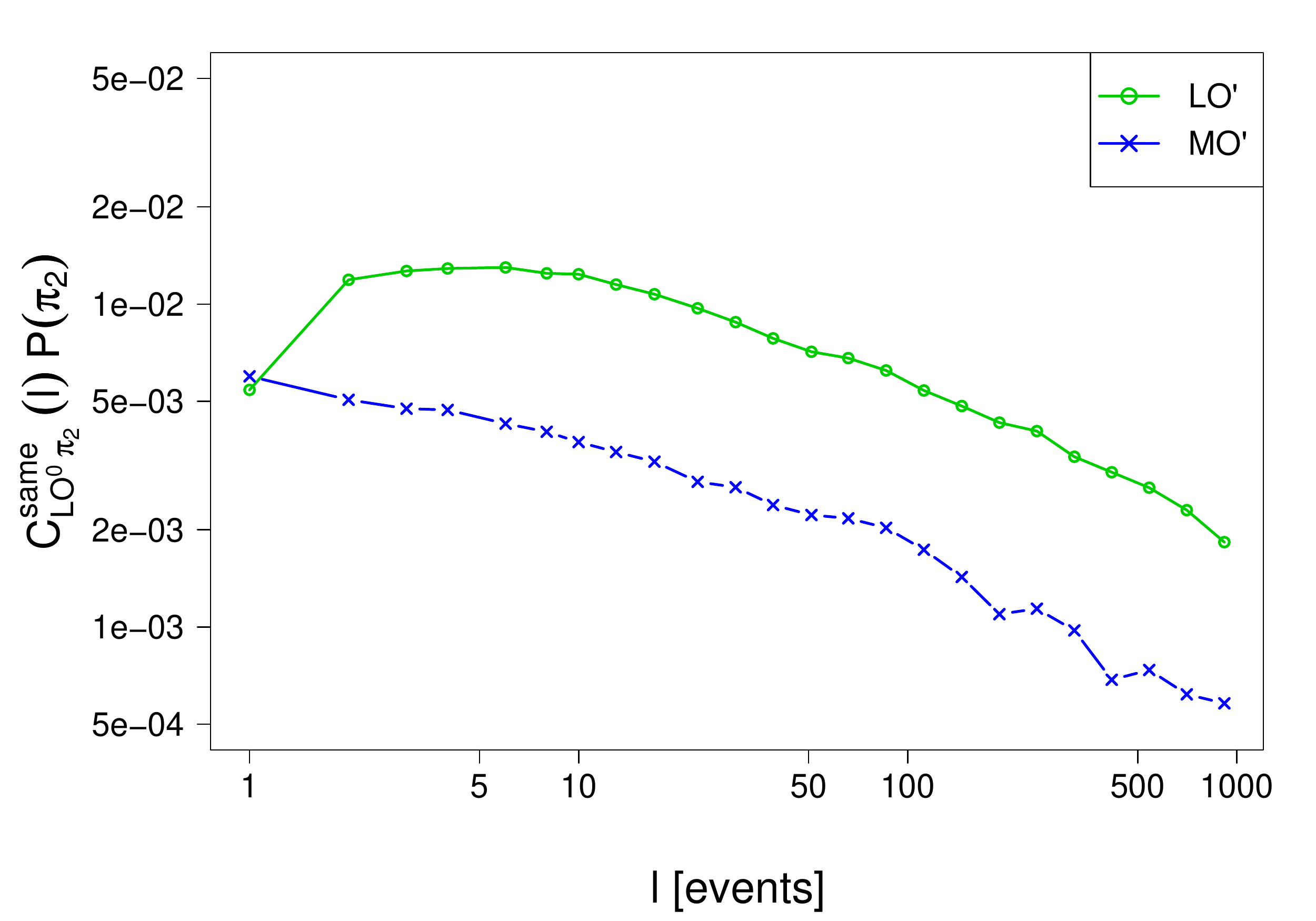}
\caption{The probability for a given broker to place price changing limit orders or market orders conditioned on starting with non-price-changing orders.  In each case we show the excess probability 
$P(\pi_2)C^{\mathrm{same}}_{\pi_1,\pi_2}(\ell)$ that the same broker $b$ issues a price-changing order with the same sign after $\ell$ trades, with $\pi_2=\LO'$ or $\MO'$. The left panel shows $\pi_1=\MOO$ and 
the right panel $\pi_1=\LOO$.}
\label{fig_corr_disaggregated}
\end{center}
\end{figure}

What transpires from these plots is that conditioned on the fact that a broker decided to execute using a non-aggressive market order $\MOO$, 
she will make roughly equal use of $\MO'$ and $\LO'$ in the future, whereas after deciding to place a non-aggressive limit order $\LOO$, the probability to continue using limit orders in the future is much larger than switching to market orders. This result, however, mostly comes from the contribution of the brokers with the smallest $f^{\MO'}_{b}$, i.e. those brokers who mostly use stealth limit orders, for which such a strategy is indeed expected.

\section{Balance between liquidity taking and liquidity providing}

\begin{figure}[h]
\begin{center}
\includegraphics[width=0.49\textwidth, angle=0]{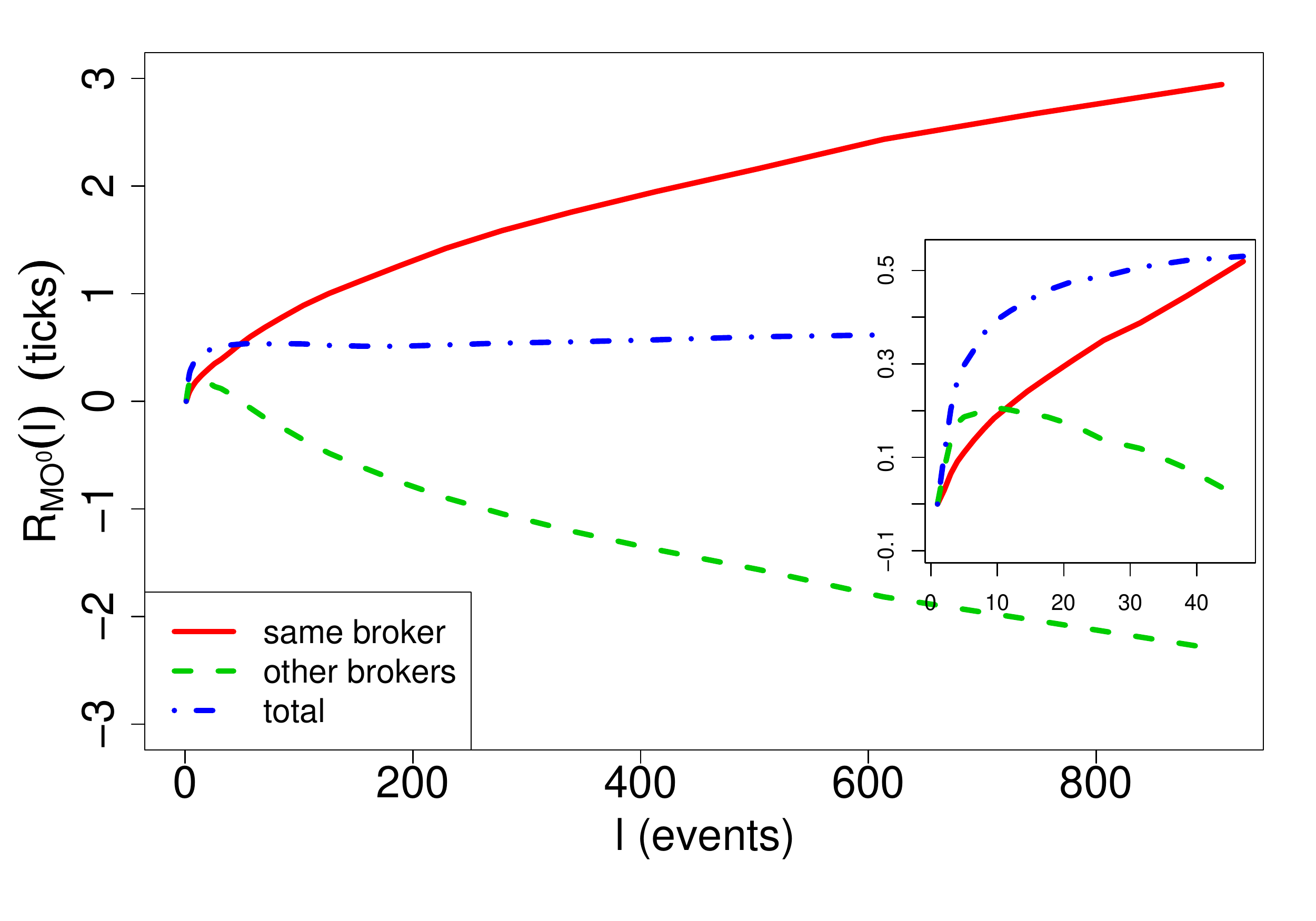}
\includegraphics[width=0.49\textwidth, angle=0]{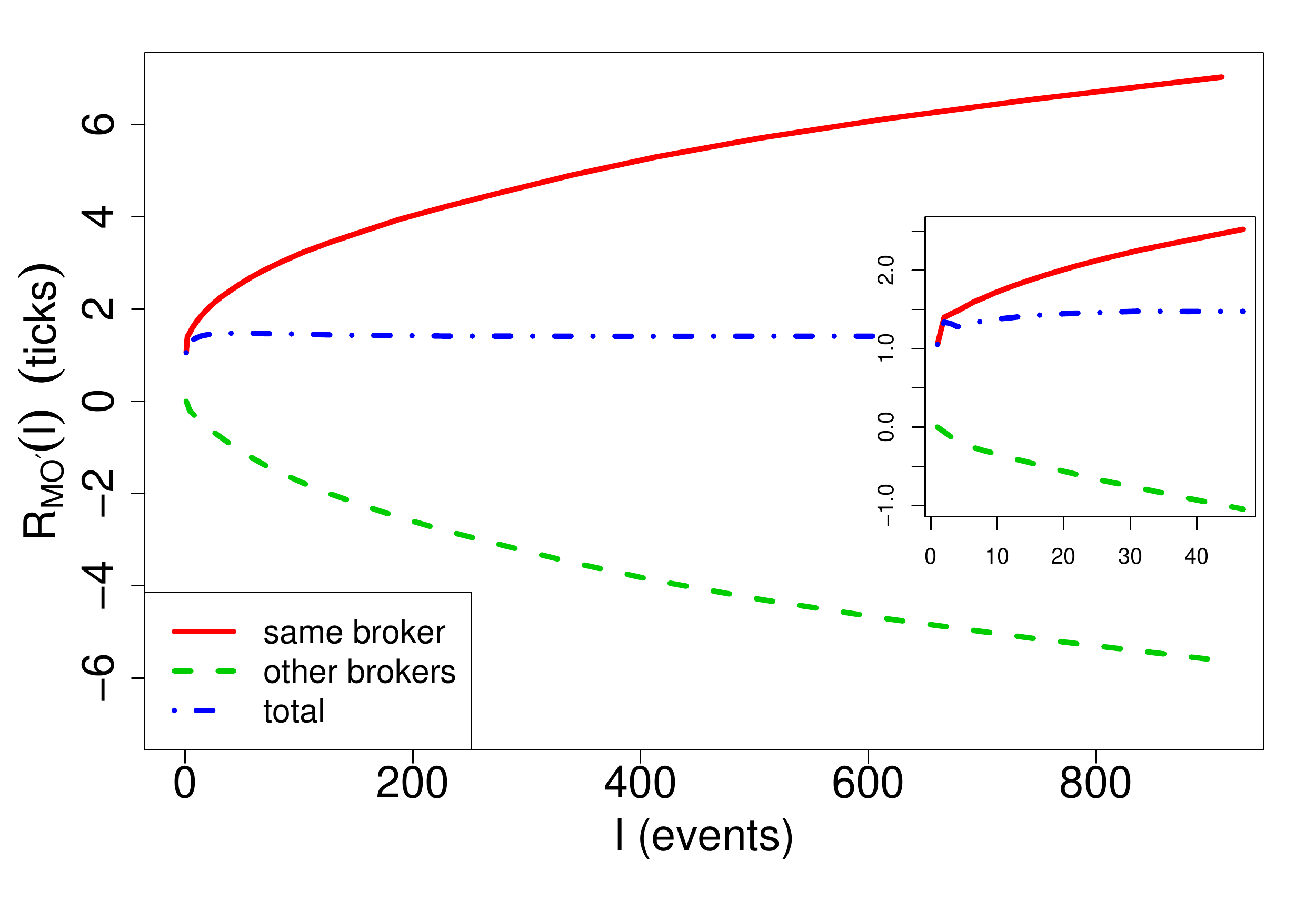}
\caption{The response function ${\cal R}_{\pi_1}(\ell)$ and its contributions coming from the orders of the same broker (${\cal R}^{\mathrm{same}}_{\pi_1}(\ell)$) 
and of different brokers (${\cal R}^{\mathrm{diff}}_{\pi_1}(\ell)$). (left) The case of $\pi_1=\MOO$. (right) The case of $\pi_1=\MO'$. The insets show a zoom for 
small $\ell$.}
\label{fig_respMO}
\end{center}
\end{figure}

\begin{figure}[h]
\begin{center}
\includegraphics[width=0.49\textwidth, angle=0]{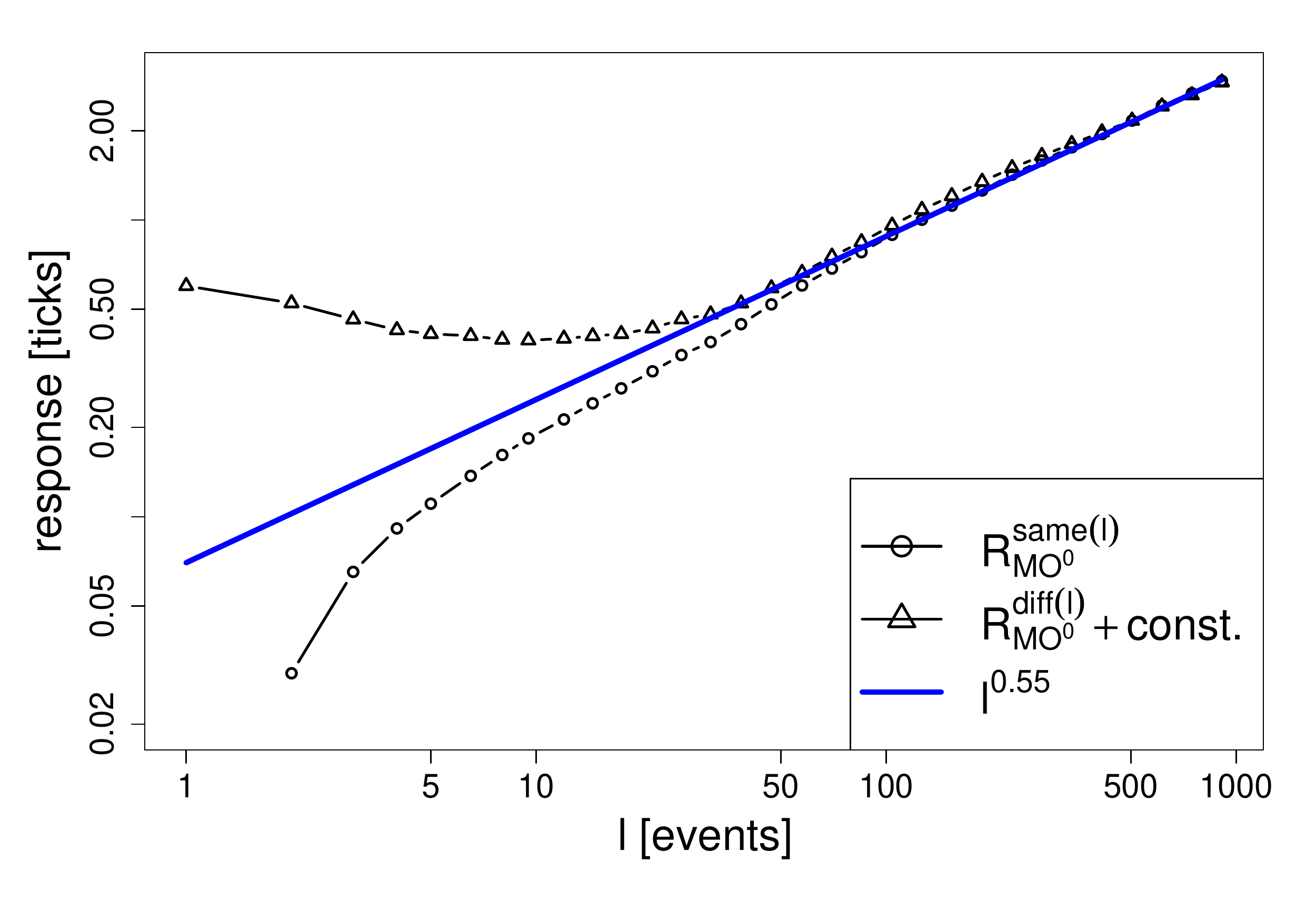}
\includegraphics[width=0.49\textwidth, angle=0]{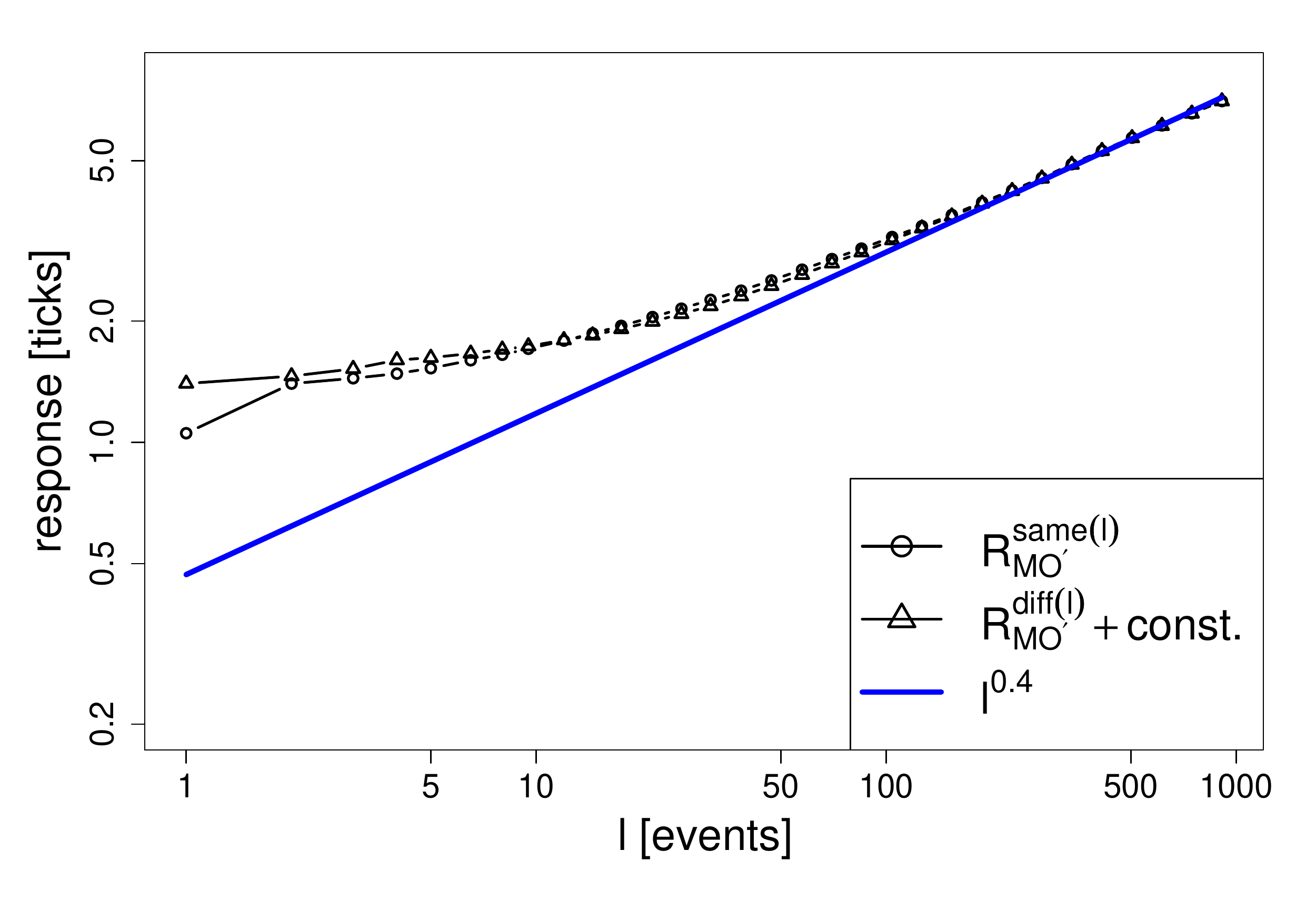}
\caption{The contributions to the response function ${\cal R}_{\pi_1}(\ell)$ form the same broker and from different brokers on a log-log scale.
We show ${\cal R}^{\mathrm{same}}_{\MO}(\ell)$, together with $-{\cal R}^{\mathrm{diff}}_{\MO}(\ell)+\mathrm{const.}$, where a constant term was just added in order to better visualise the 
similarity of the curves. (left) The case of $\pi_1=\MOO$. (right) The case of $\pi_1=\MO'$.}
\label{fig_respMO_log}
\end{center}
\end{figure}

Let us investigate another aspect of the intertwined liquidity dynamics, and present the most striking result of our study. It is known from previous results \cite{hasbrouck1991, bouchaud2004, Eisler2010} that 
the average impact of market orders, ${\cal R}_{\MO}(\ell)$, first increases rapidly with $\ell$ and then becomes flat for large $\ell$. To understand the reason 
for the flattening of the response function, we studied the contributions coming from the actions of the same broker and of other brokers. When disaggregating these two effects, we find that the flat response 
function comes from a nearly exact cancellation between ${\cal R}_{\MO}^{\mathrm{same}}(\ell)$, which increases as roughly $\sqrt{\ell}$, and 
${\cal R}_{\MO}^{\mathrm{diff}}(\ell)$, which decreases as roughly $\sqrt{\ell}$. The growth of 
${\cal R}^{\mathrm{same}}_{\MO}(\ell)$ can be directly understood from the self-correlation $C^{\mathrm{same}}(\ell)$ described above\footnote{More 
precisely, for large tick stocks the response is related to the integral of the correlation functions \cite{Eisler2010}. Therefore, since the correlations 
$C^{\mathrm{same}}_{\MO,\pi_2}(t')$ all have a power law decay with exponents close to $\gamma=0.5$, we expect the response to increase with an exponent close to 
$1-\gamma=0.5$. In fact ${\cal R}^{\mathrm{same}}_{\MOO}(\ell)$ increases as a power law with an exponent between $0.5$ and $0.63$ for all the studied stocks, while ${\cal R}^{\mathrm{same}}_{\MO'}(\ell)$ show 
slightly lower exponents between $0.38$ and $0.51$.}. 
The impact functions illustrating this behaviour for event types $\MOO$ and $\MO'$ can be seen in Figure \ref{fig_respMO}. 
To better show the power law increase of the absolute values, in Figure \ref{fig_respMO_log} we plot ${\cal R}^{\mathrm{same}}_{\MO}(\ell)$ on a log-log scale, together with 
$-{\cal R}^{\mathrm{diff}}_{\MO}(\ell)+\mathrm{const.}$, where a constant term was just added in order to better visualise the 
similarity of the curves. 

Interestingly, the sum of these two responses give a total response that is much weaker in absolute value and is flat for time lags $\ell\gtrapprox100$. This is 
the central result of our study. The short time behaviour ($\ell\lessapprox10-20$) is different when  (i) the initial order left the price unchanged ($\MOO$) 
and (ii) when it changed the price ($\MO'$). In case (i) the contribution coming from different brokers is initially positive and then becomes negative, while 
in case (ii) it is immediately negative. Therefore, upon an aggressive buy market order ($\MO'$) from one broker, the other brokers (probably those with small 
$f^{\MO'}_{b}$'s) react by immediately providing liquidity to the market, and continue to do so during the whole `buying spree', thereby limiting the total upward 
price shift. After a non-aggressive market order, on the other hand, the herding mechanism described in the previous section explains the initial positive 
contribution of ${\cal R}_{\MOO}^{\mathrm{diff}}(\ell)$ seen in Figure \ref{fig_respMO}, left (inset).

The above findings extend to the decomposition of all types of impact functions ${\cal R}_{\pi}(\ell)$.
After any type of event, the response due to the same broker's actions is monotonically increasing. In contrast, the response due to the rest of the market is 
always monotonically decreasing for $\ell\gtrapprox10-20$~\footnote{The frequency of events varies among stocks and changes in time. On average the frequency 
of events in the period studied was 0.28 events/sec.}.

\section{Conclusions}
\label{concl}

The aim of this paper was to use the formalism introduced in \cite{Eisler2010} to  exploit a database where the broker initiating an order book event can be identified, and decompose the correlations and 
response functions into different contributions.  This allowed us to identify several interesting regularities in order placement, as summarized below:\

\begin{itemize}
\item In Section 3 we present clear evidence that brokers are heterogeneous in their liquidity provision.  Different brokers use different type of strategies, from liquidity providing strategies with a small 
fraction of market orders to liquidity taking strategies with a very large fraction of market orders.
\item In Section 4 we confirm that the long-range correlation in the sign of market orders comes mostly from order splitting from a unique broker.  There is, however, a certain degree of herding behaviour 
from other brokers which is evident as long as the price does not change. After a price changing market order, however, the broker responsible for it continues trading in the same direction regardless of 
his own impact, whereas other brokers start firing market orders in the opposite direction.
\item The central finding of our study is that the total impact of market orders results from the nearly perfect compensation between two opposite contributions: one resulting from the accumulation of orders in the same direction coming from 
the same broker, while the reaction of brokers who provide liquidity results in an impact of roughly equal magnitude but opposite sign.  (The contribution by other brokers is always a bit smaller, so that 
the total impact has the correct sign).
\end{itemize}

These results suggest a picture of the ecology of markets anticipated in several papers (see e.g. \cite{handa1998,farmer2002,bouchaud2006,lillo2008,moro2009}), where agents are both broadly heterogeneous 
in their expectations and 
strategies, and strongly interacting, with a complex intertwined dynamics between liquidity providers and liquidity takers. It is tempting to conjecture that these ingredients are crucial to understand
the incipient instabilities of financial markets, epitomized by price jumps and volatility clustering \cite{joulin2008,bouchaud2011,kirilenko2010}.  

\section*{Acknowledgements}
FL acknowledges partial support from the grant SNS11LILLB ``Price formation, agentÕs heterogeneity, and market efficiency''.

\appendix

\section{Significance test for the impact of different groups of brokers}
\label{app}

We used Student's t-test on the ensemble of the immediate price impacts both to value the hypothesis that the impact of $MO'$s and $LO'$s are different (blue and green curves in Fig.~(\ref{fig_brokers}) right panel) and to test if the slope of the curves are different from zero. The hypothesis that the immediate impacts in Fig~(\ref{fig_brokers}) come from distributions with equal means can be rejected at the 5\% level for 41 pairs out of the possible 45 pairs (see Table). This tells us that large part of the differences between the points are significant.

\begin{table}[h!]
\begin{center}
{\footnotesize
\begin{tabular}{| c | c | c | c | c | c | c | c | c | c | c |}
\hline
&Q1-$LO'$ &Q2-$LO'$ & Q3-$LO'$ & Q4-$LO'$ & Q5-$LO'$ & Q1-$MO'$ & Q2-$MO'$ & Q3-$MO'$ & Q4-$MO'$ & Q5-$MO'$\\
\hline
Q1-$LO'$ & 1 & 0.00* & 0.00* & 0.00* & 0.00* & 0.00* & 0.00* & 0.00* & 0.00* & 0.00*\\
Q2-$LO'$ & 0.00* & 1 & 0.00* & 0.00* & {\bf 0.11} & 0.00* & 0.00* & 0.00* & 0.00* & 0.00* \\
Q3-$LO'$ & 0.00* & 0.00* & 1 & {\bf 0.99} & 0.00* & 0.00* & 0.00* & 0.00* & 0.02** & 0.01** \\
Q4-$LO'$ & 0.00* & 0.00* & {\bf 0.99} & 1 & 0.00* & 0.00* & 0.00* & 0.00* & {\bf 0.08} & {\bf 0.05} \\
Q5-$LO'$ & 0.00* & {\bf 0.11} & 0.00* & 0.00* & 1 & 0.00* & 0.00* & 0.00* & 0.00* & 0.00* \\
Q1-$MO'$ & 0.00* & 0.00* & 0.00* & 0.00* & 0.00* & 0.00* & 0.04** & 0.00* & 0.00* & 0.00* \\
Q2-$MO'$ & 0.00* & 0.00* & 0.00* & 0.00* & 0.00* & 0.04**  & 1 & 0.03**  & 0.00* & 0.00* \\
Q3-$MO'$ & 0.00* & 0.00* & 0.00* & 0.00* & 0.00* & 0.00* & 0.03** & 1 & 0.00* & 0.00* \\
Q4-$MO'$ & 0.00* & 0.00* & 0.02** & {\bf 0.08} & 0.00* & 0.00* & 0.00* & 0.00* & 1 & 0.00* \\
Q5-$MO'$ & 0.00* & 0.00* & 0.01** & {\bf 0.05} & 0.00* & 0.00* & 0.00* & 0.00* & 0.00* & 1 \\
\hline
\end{tabular}
}
\end{center}
\caption{\label{table2} The p-values of the hypothesis test that the points presented in Fig.~(\ref{fig_brokers}) come from distributions with equal means. Out of the 45 possible point pairs (the matrix is symmetric) in 41 cases the hypothesis can be rejected at the 5\% significance level and in 37 cases the hypothesis can be rejected at the 1\% significance level. (* denotes significance at the 1\% level, ** denotes significance at the 5\% level.) Q1,\dots,Q5 denote the 5 quantiles in Fig~(\ref{fig_brokers}) with increasing $f_b^{MO'}$.}
\end{table}

\section{Plots of all the correlation functions}
In this appendix we show all the 72 possible correlation functions mentioned in Section 4.
\label{more_plots}
\begin{figure}[h]
\begin{center}
\includegraphics[width=0.49\textwidth]{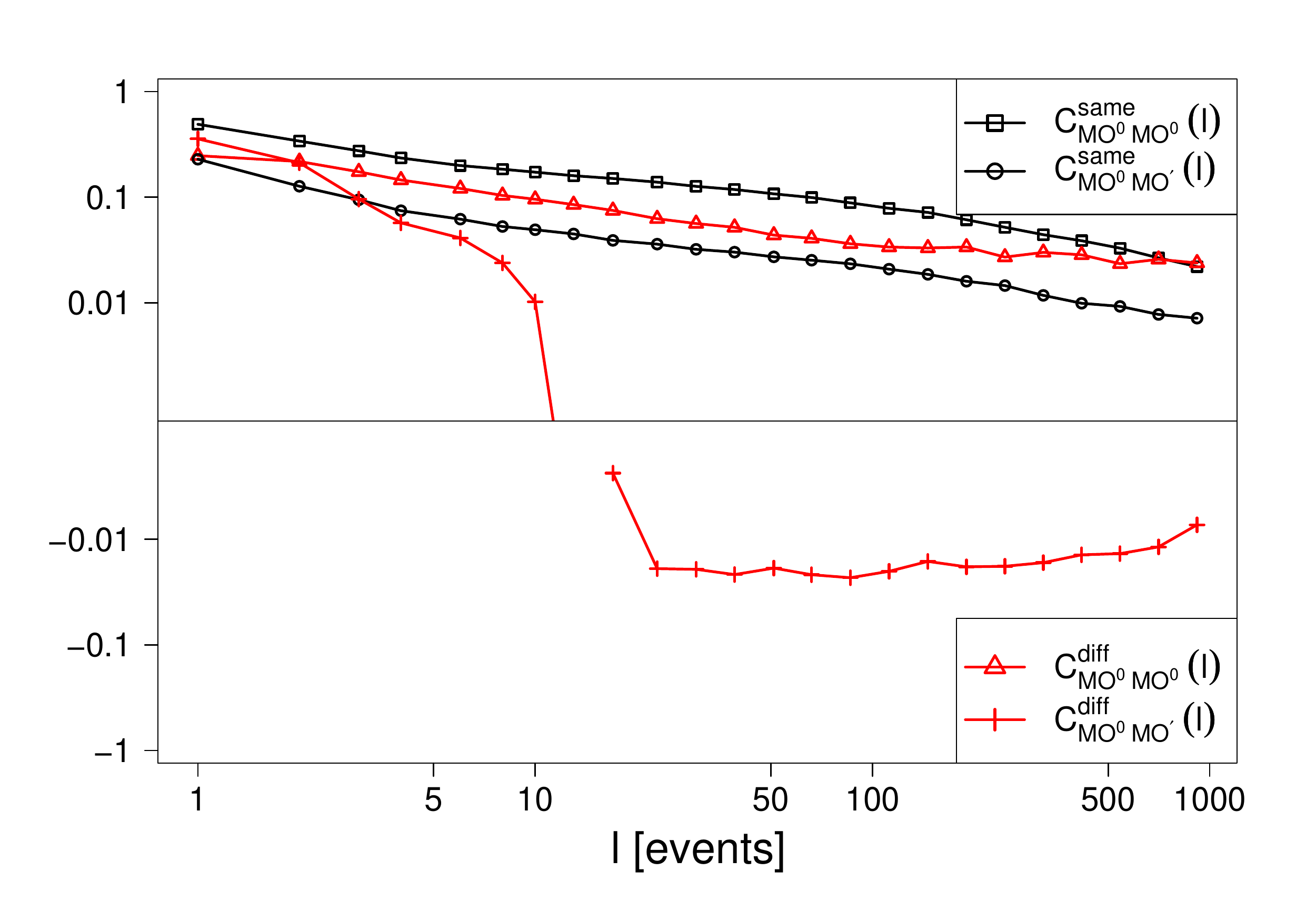}
\includegraphics[width=0.49\textwidth]{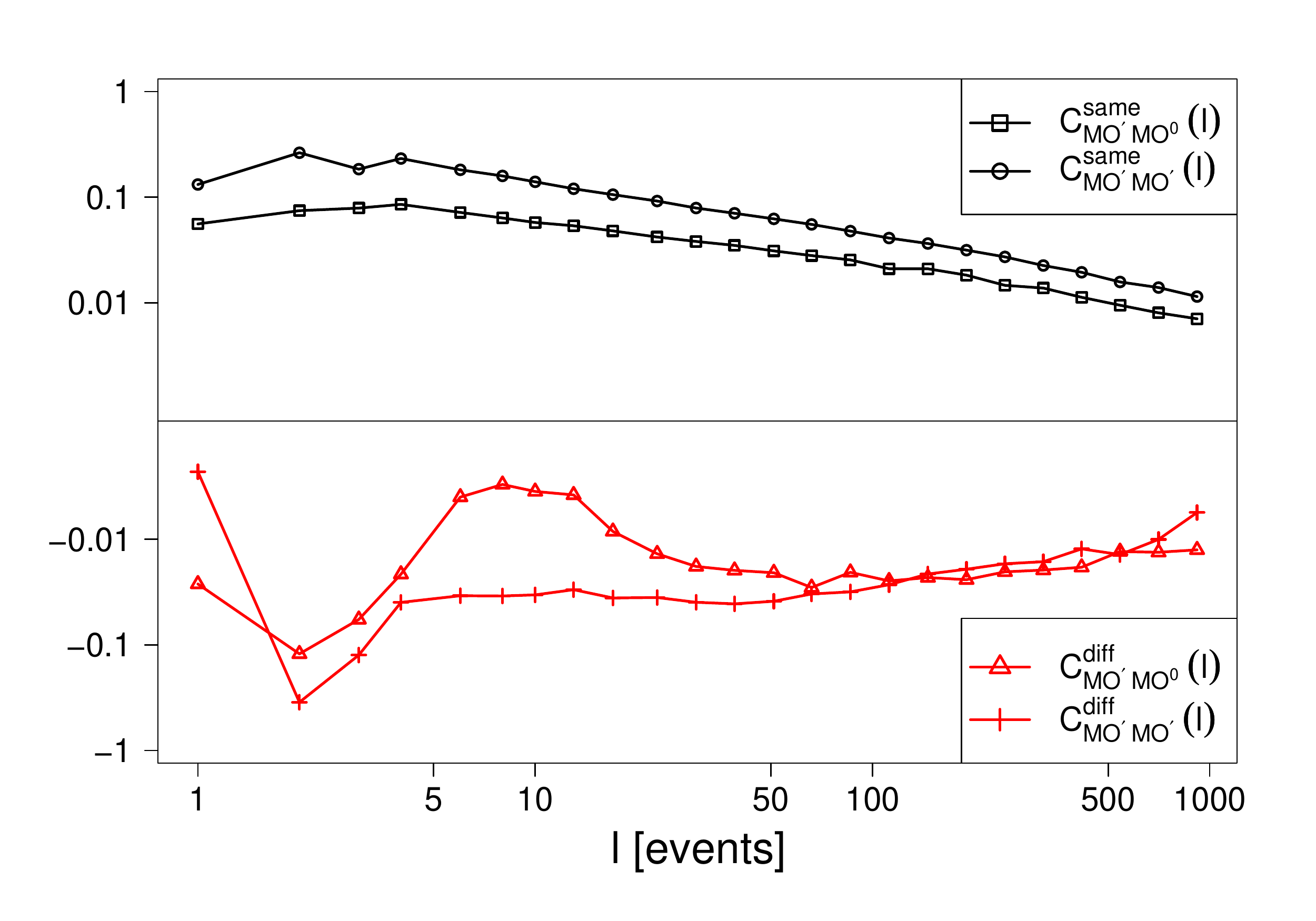}
\caption{Plots of the correlation $C_{\pi_1,\MO}(\ell)$. {\it (left)} $\pi_1=\MOO$; {\it (right)} $\pi_1=\MO'$}
\label{fig1}
\end{center}
\end{figure}

\begin{figure}[h]
\begin{center}
\includegraphics[width=0.49\textwidth]{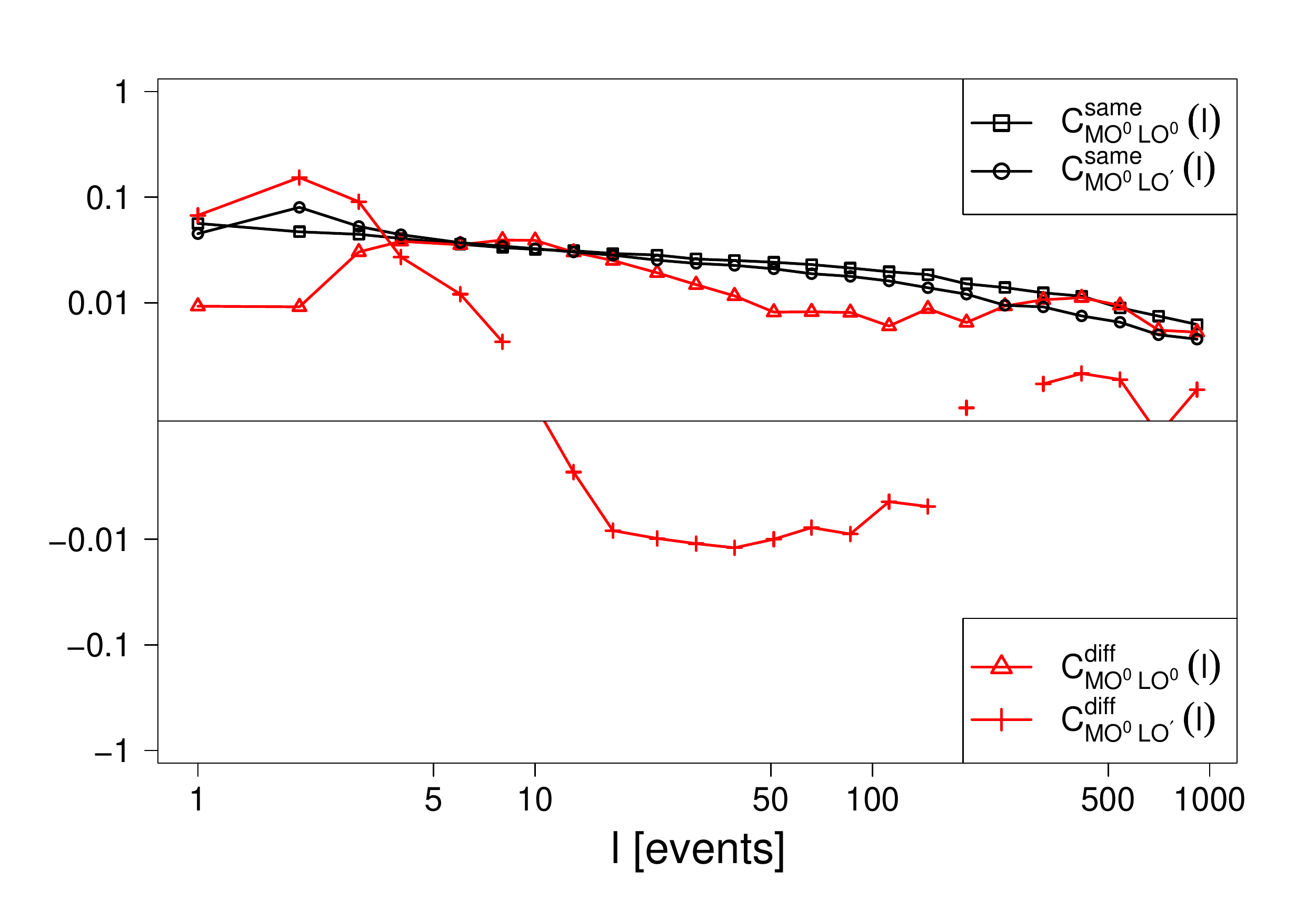}
\includegraphics[width=0.49\textwidth]{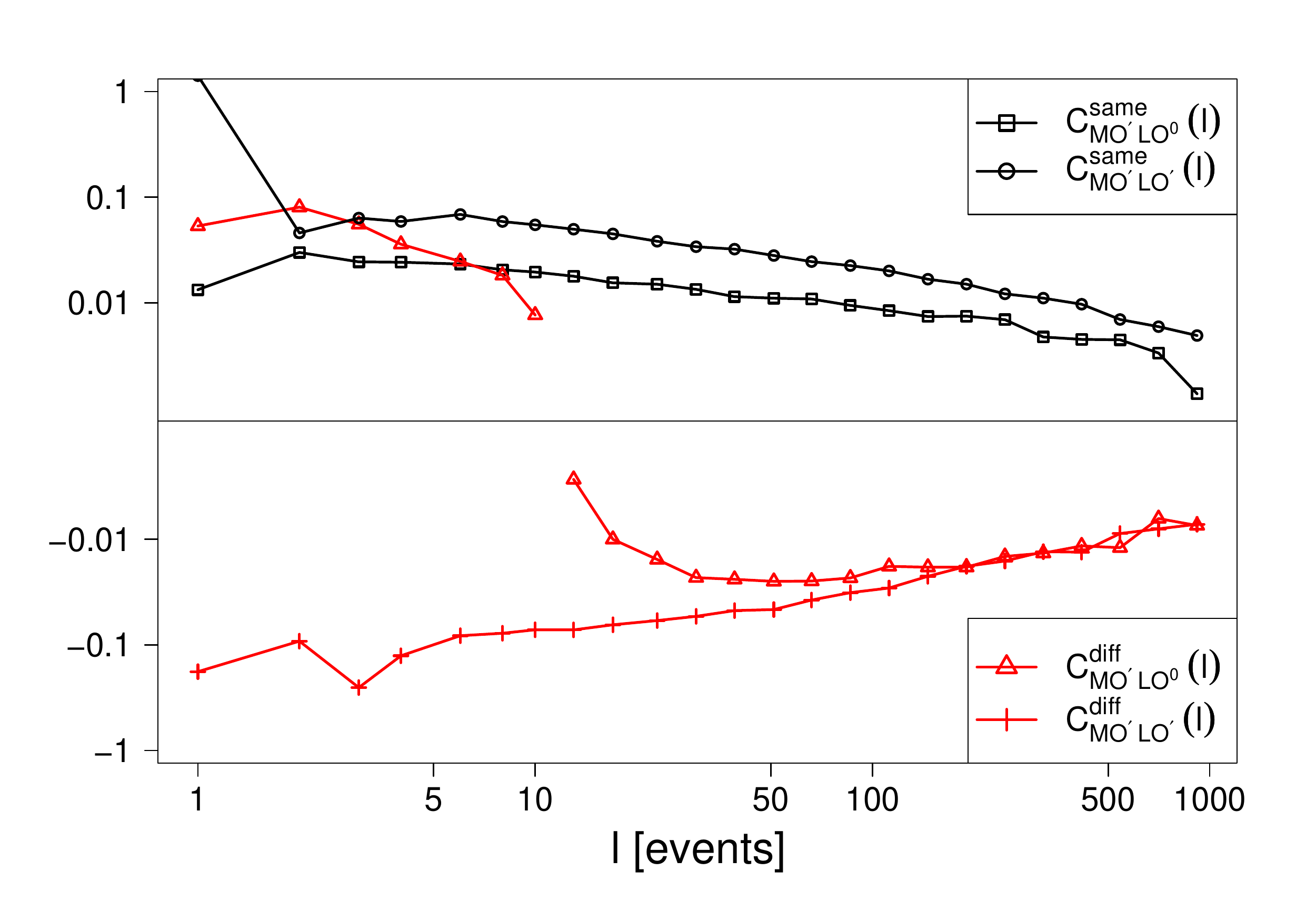}
\caption{Plots of the correlation $C_{\pi_1,\LO}(\ell)$. {\it (left)} $\pi_1=\MOO$; {\it (right)} $\pi_1=\MO'$}
\label{fig2}
\end{center}
\end{figure}

\begin{figure}[h]
\begin{center}
\includegraphics[width=0.49\textwidth]{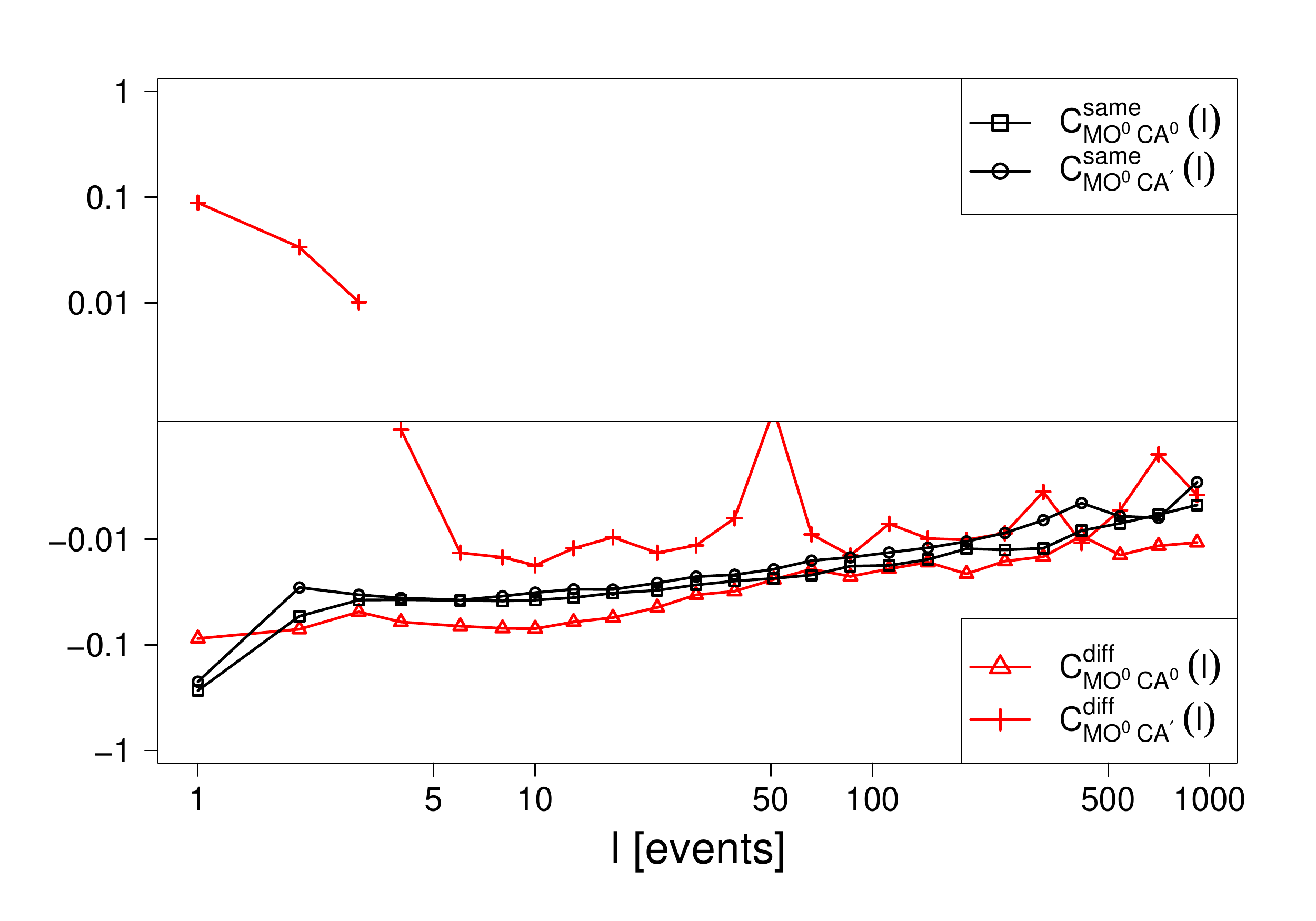}
\includegraphics[width=0.49\textwidth]{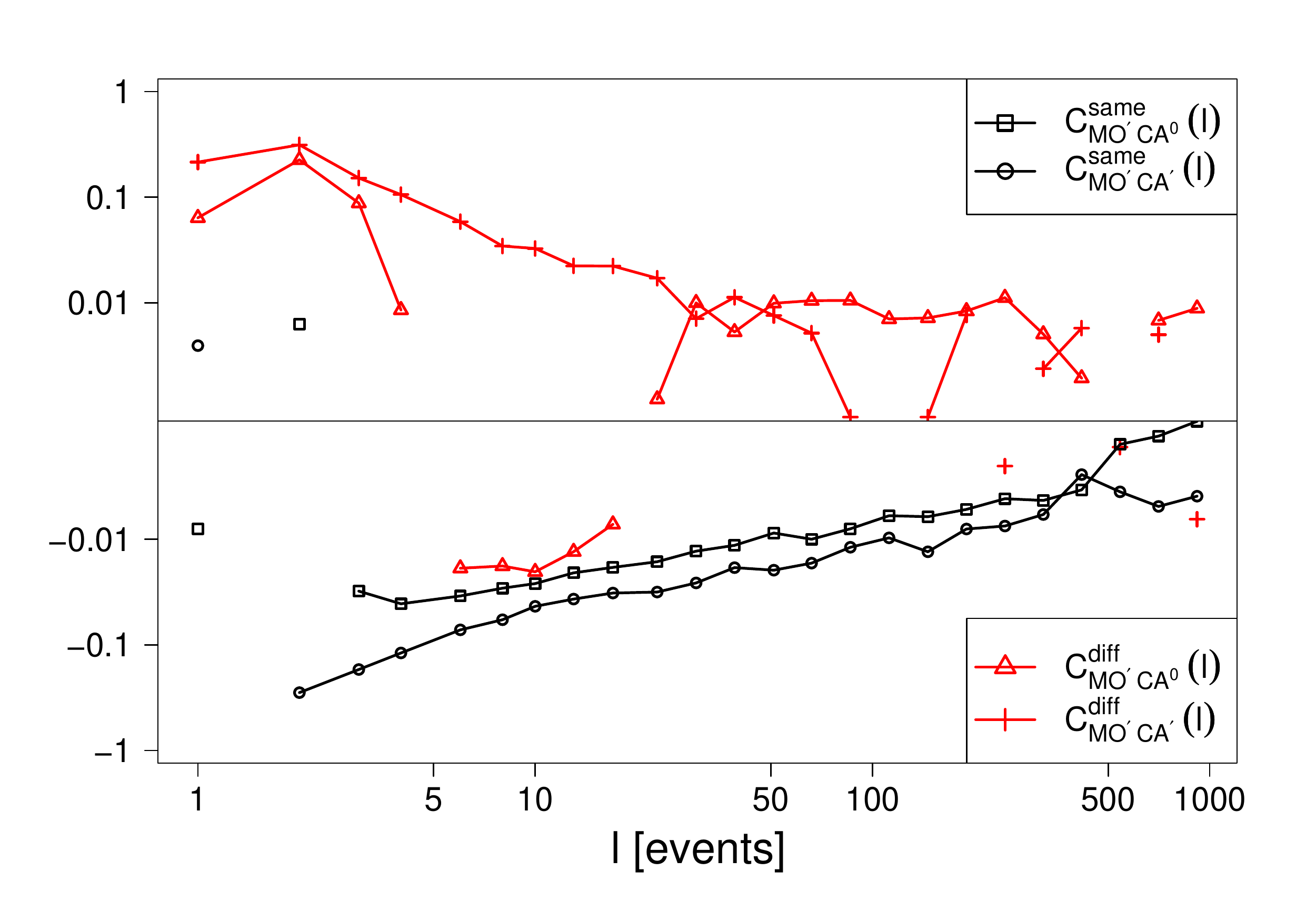}
\caption{Plots of the correlation $C_{\pi_1,\CA}(\ell)$. {\it (left)} $\pi_1=\MOO$; {\it (right)} $\pi_1=\MO'$}
\label{fig3}
\end{center}
\end{figure}

\begin{figure}[h]
\begin{center}
\includegraphics[width=0.49\textwidth]{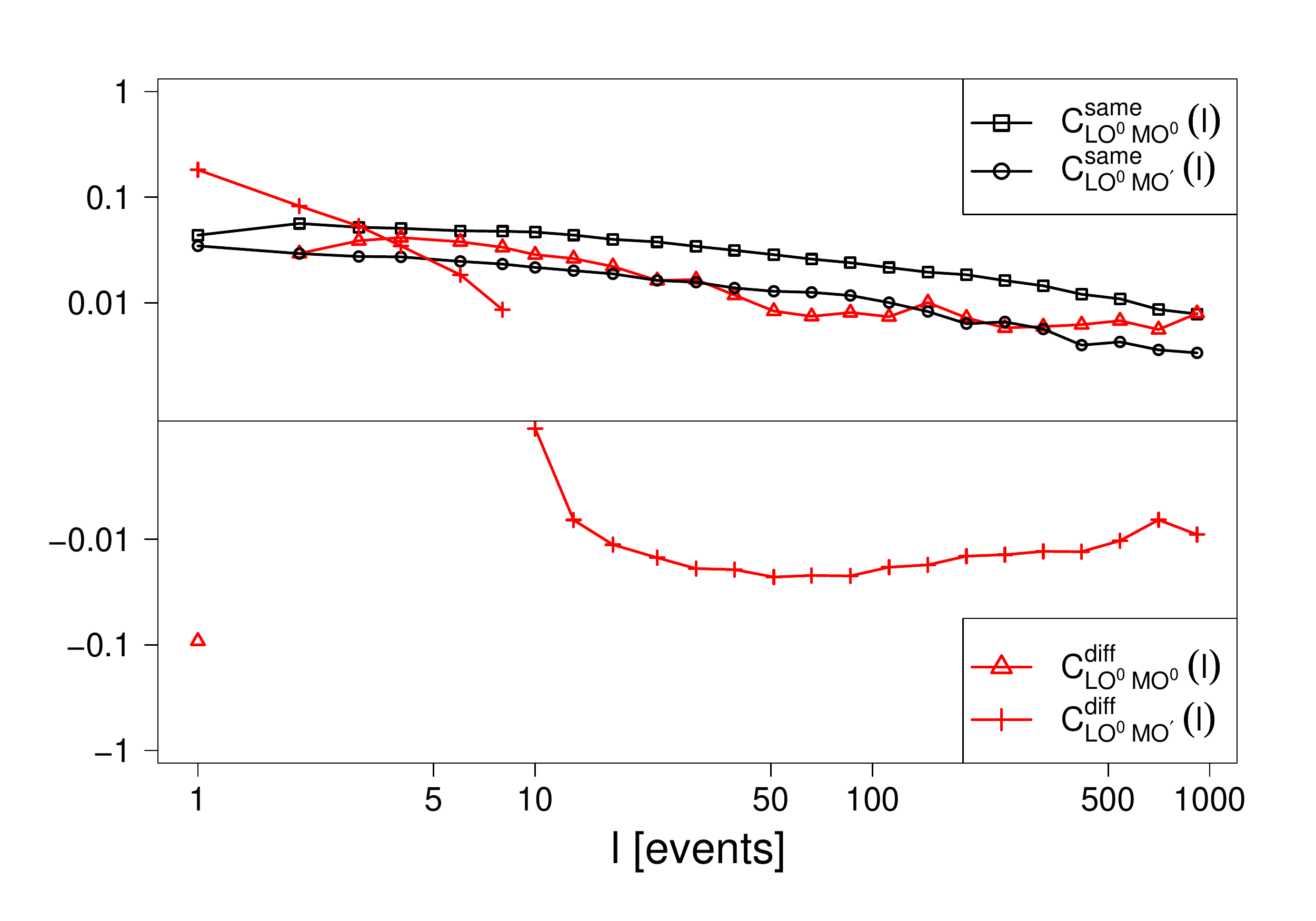}
\includegraphics[width=0.49\textwidth]{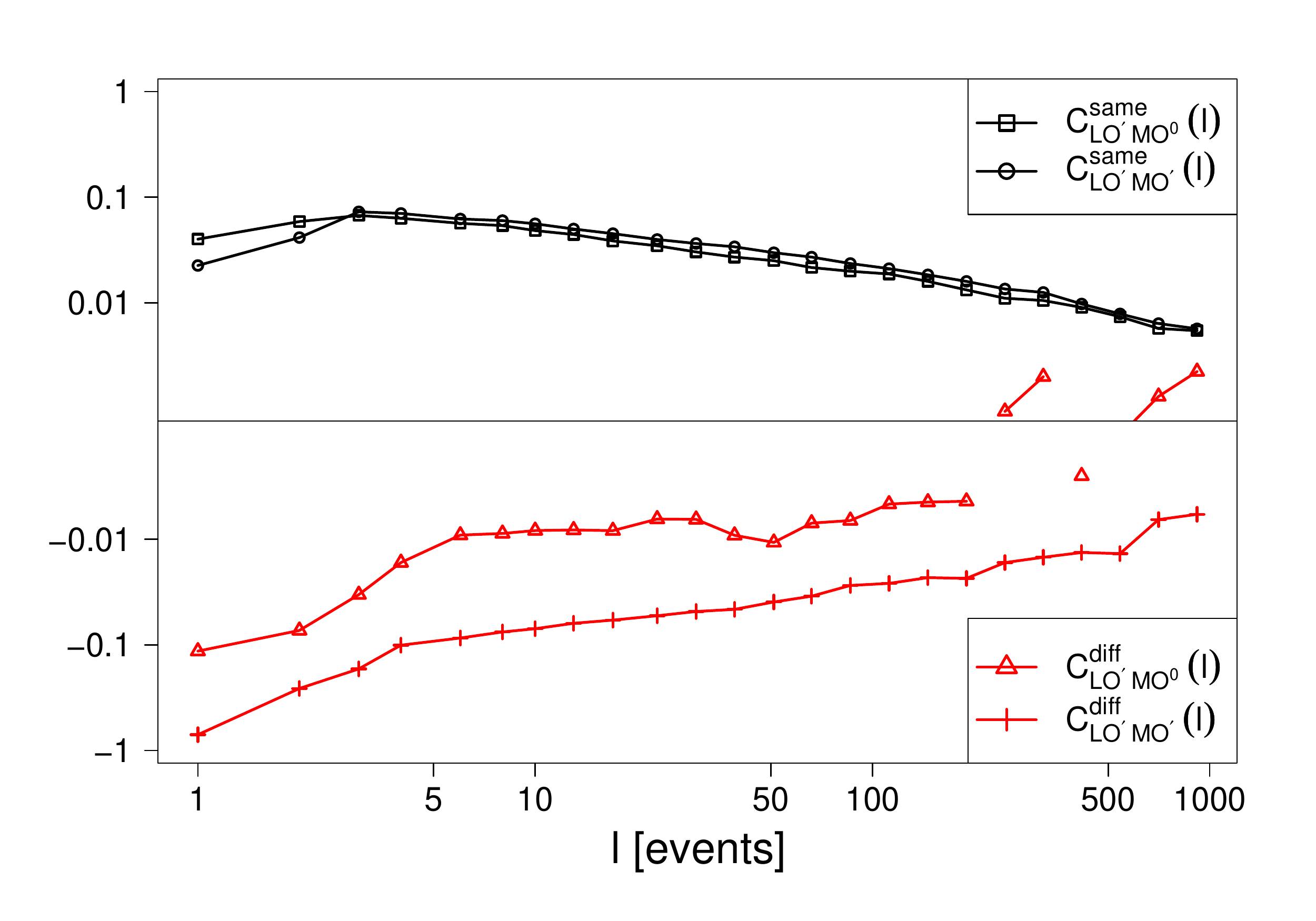}
\caption{Plots of the correlation $C_{\pi_1,\MO}(\ell)$. {\it (left)} $\pi_1=\LOO$; {\it (right)} $\pi_1=\LO'$}
\label{fig4}
\end{center}
\end{figure}

\begin{figure}[h]
\begin{center}
\includegraphics[width=0.49\textwidth]{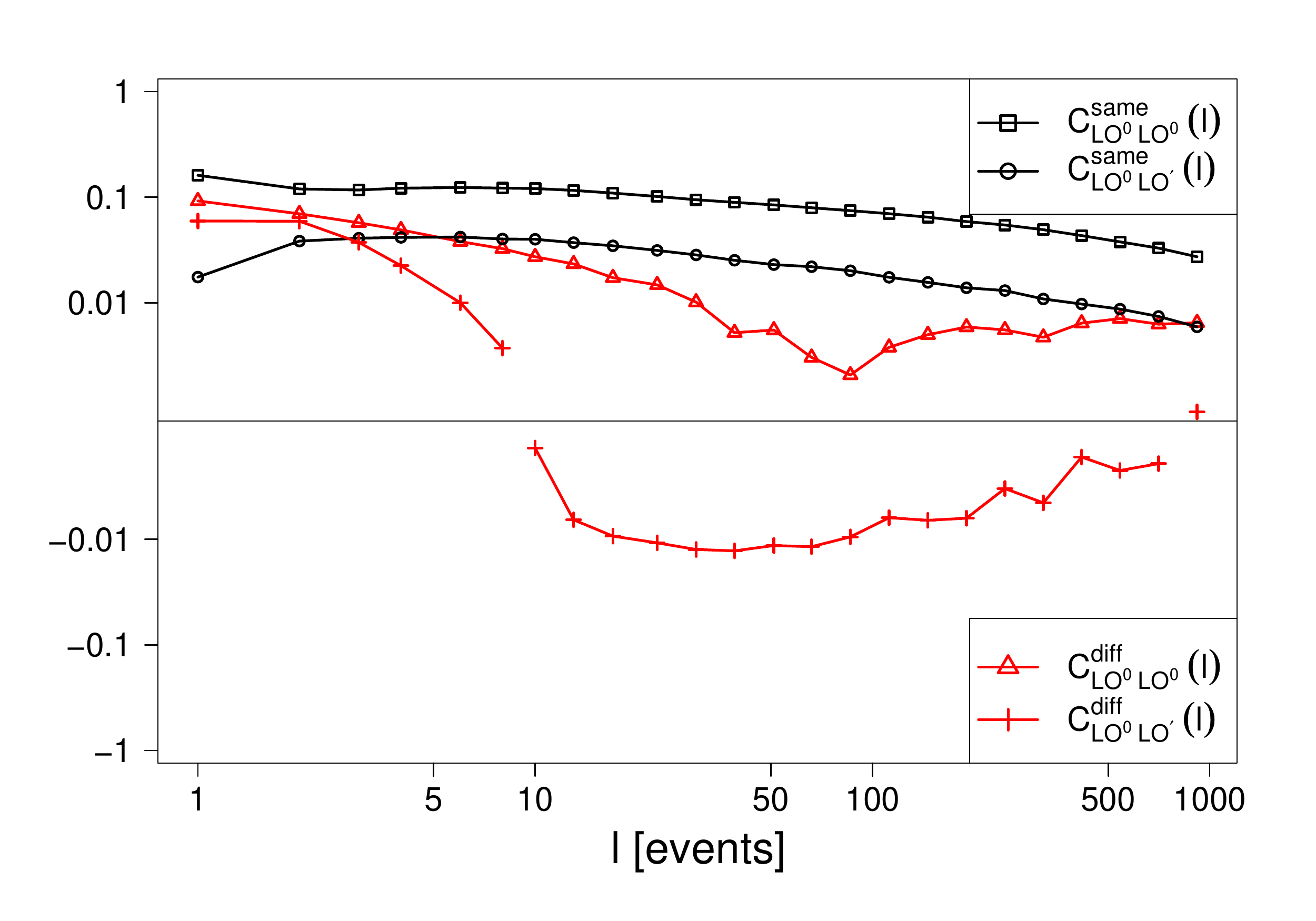}
\includegraphics[width=0.49\textwidth]{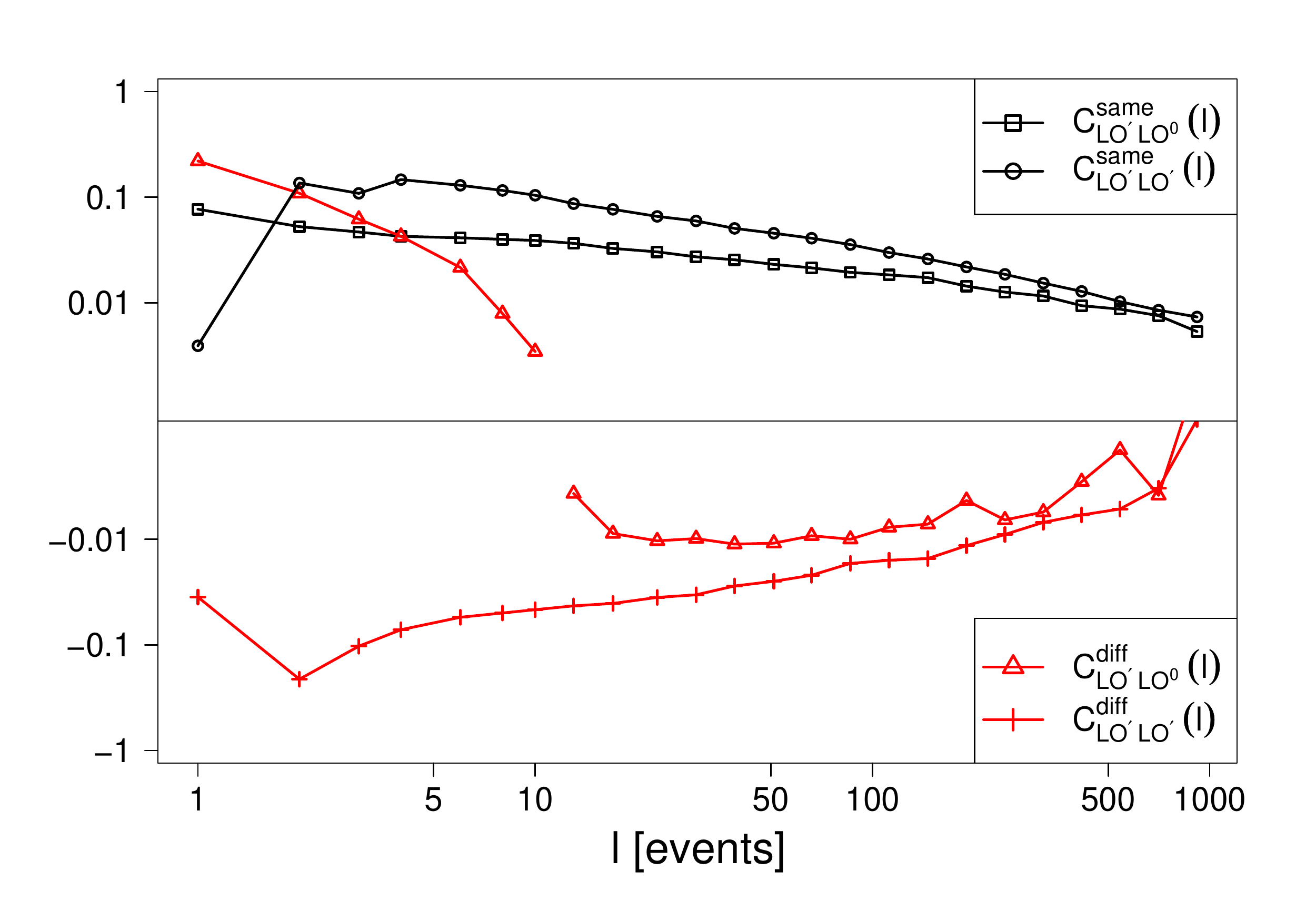}
\caption{Plots of the correlation $C_{\pi_1,\LO}(\ell)$. {\it (left)} $\pi_1=\LOO$; {\it (right)} $\pi_1=\LO'$}
\label{fig5}
\end{center}
\end{figure}

\begin{figure}[h]
\begin{center}
\includegraphics[width=0.49\textwidth]{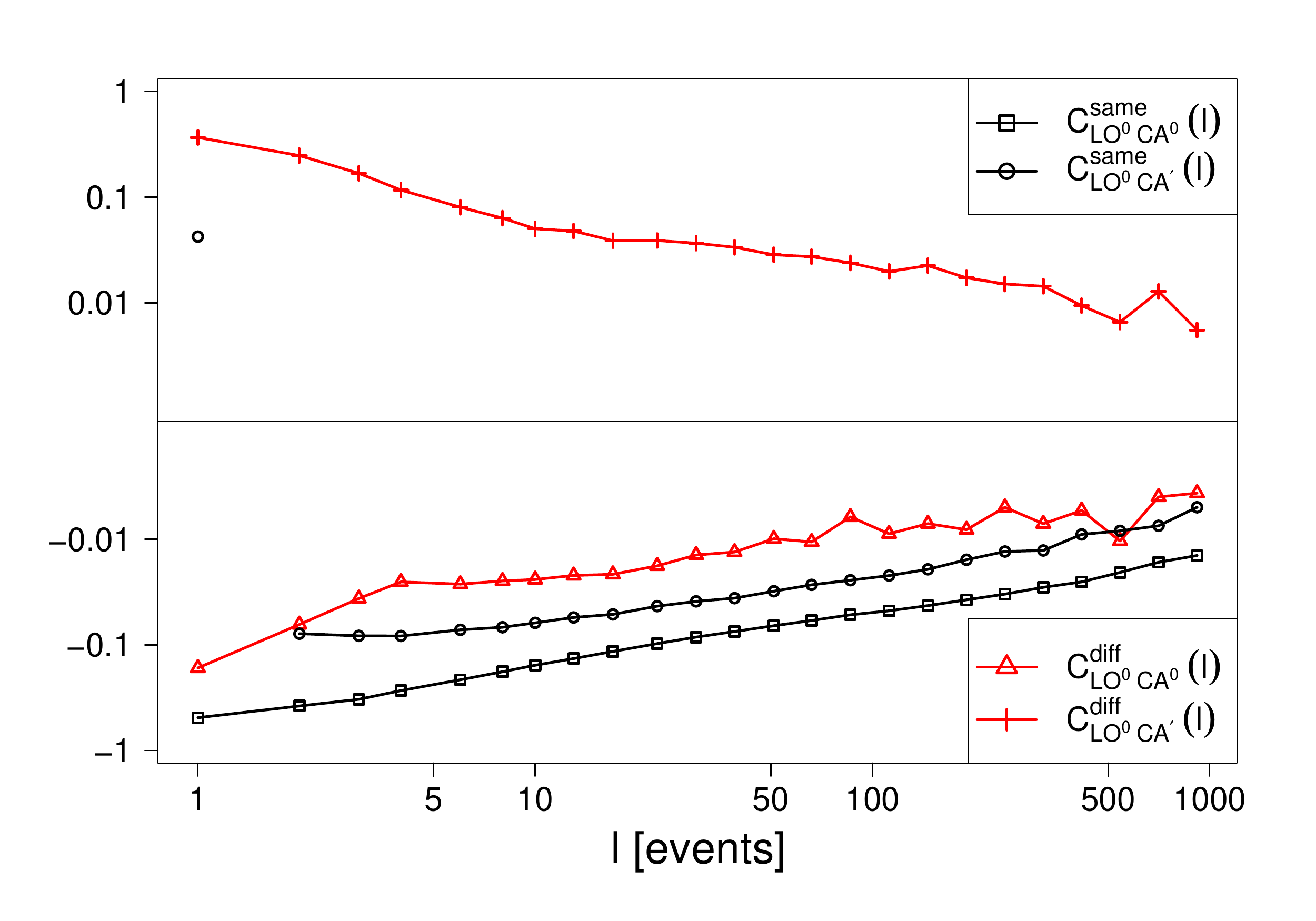}
\includegraphics[width=0.49\textwidth]{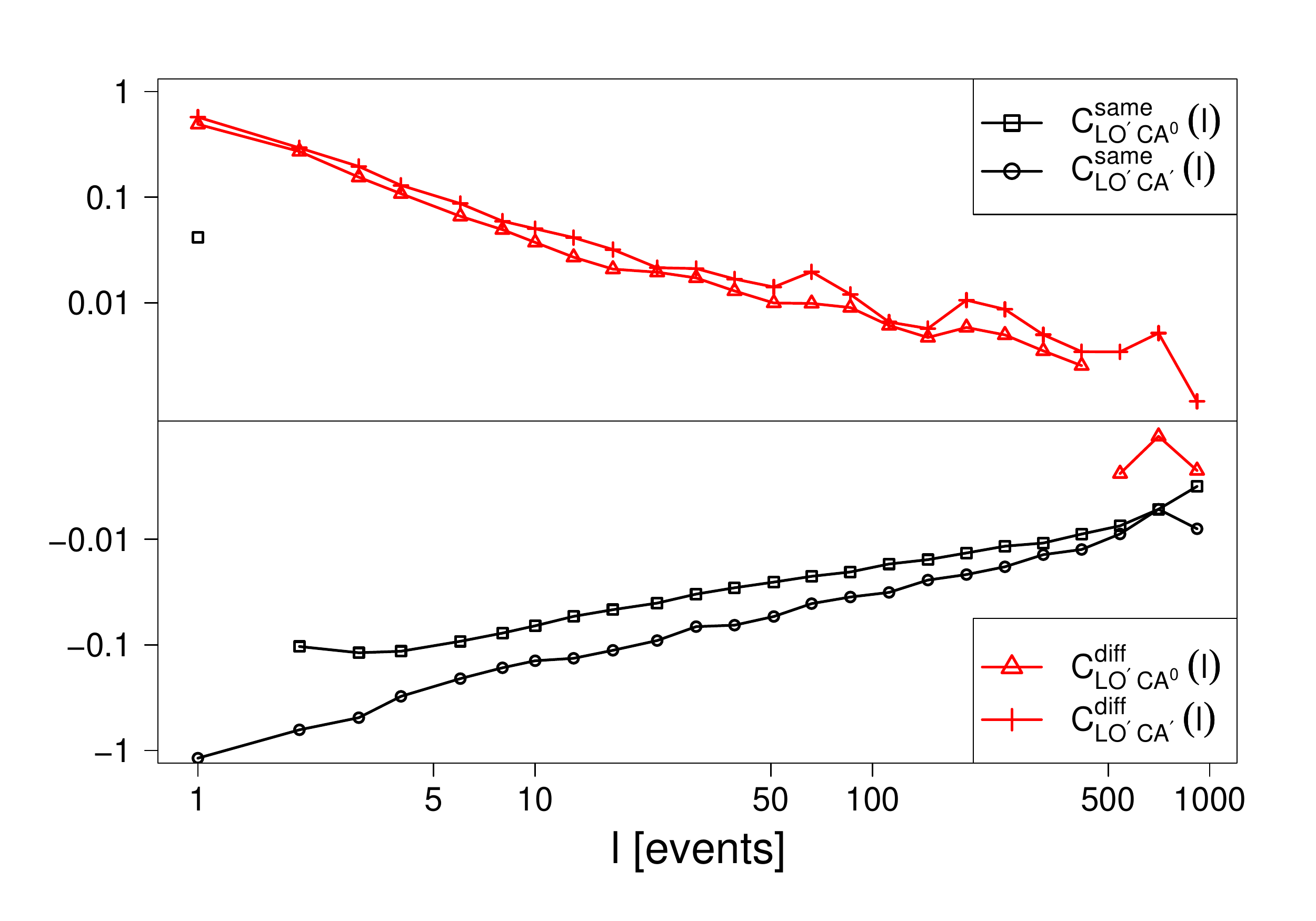}
\caption{Plots of the correlation $C_{\pi_1,\CA}(\ell)$. {\it (left)} $\pi_1=\LOO$; {\it (right)} $\pi_1=\LO'$}
\label{fig6}
\end{center}
\end{figure}

\begin{figure}[h]
\begin{center}
\includegraphics[width=0.49\textwidth]{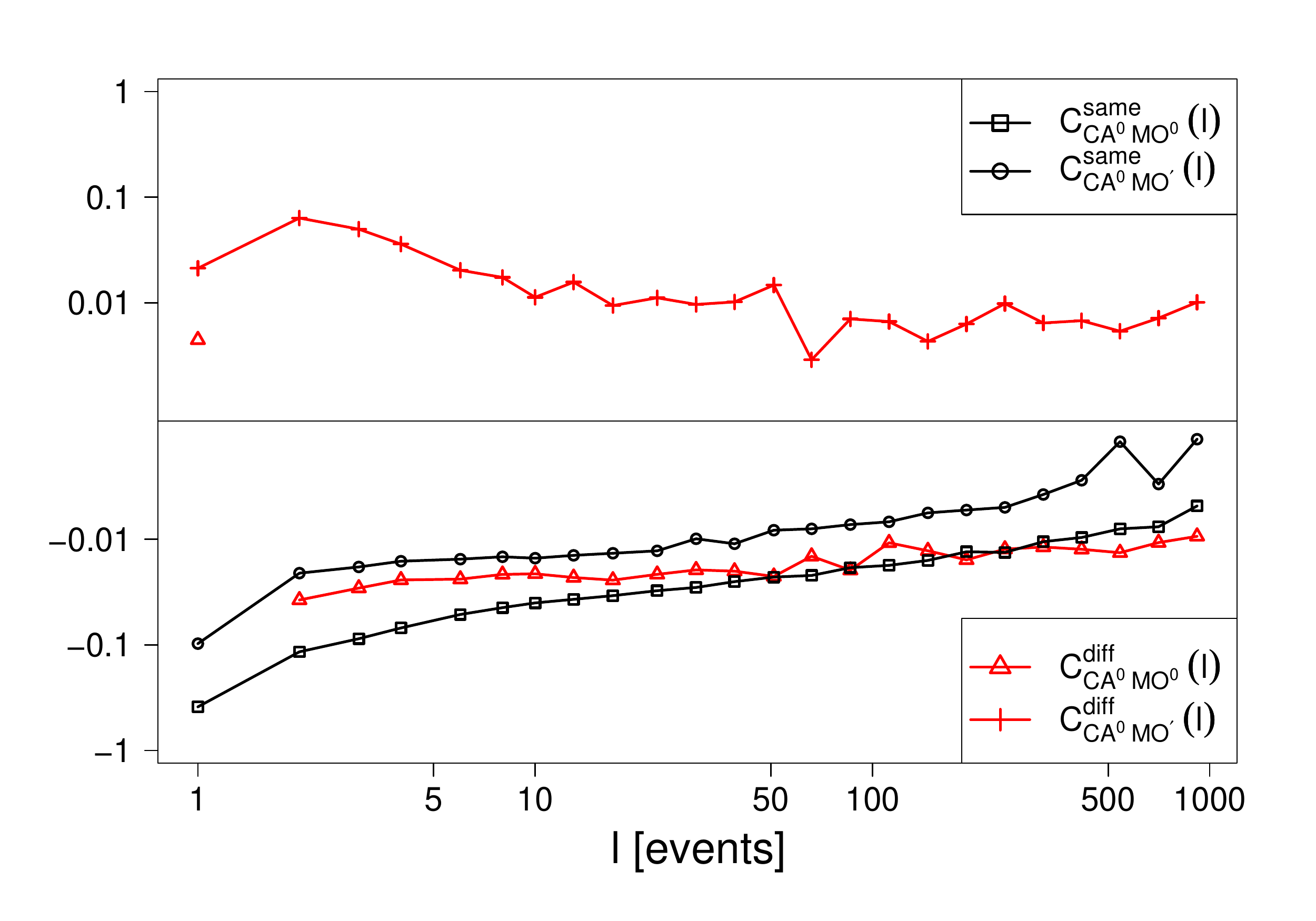}
\includegraphics[width=0.49\textwidth]{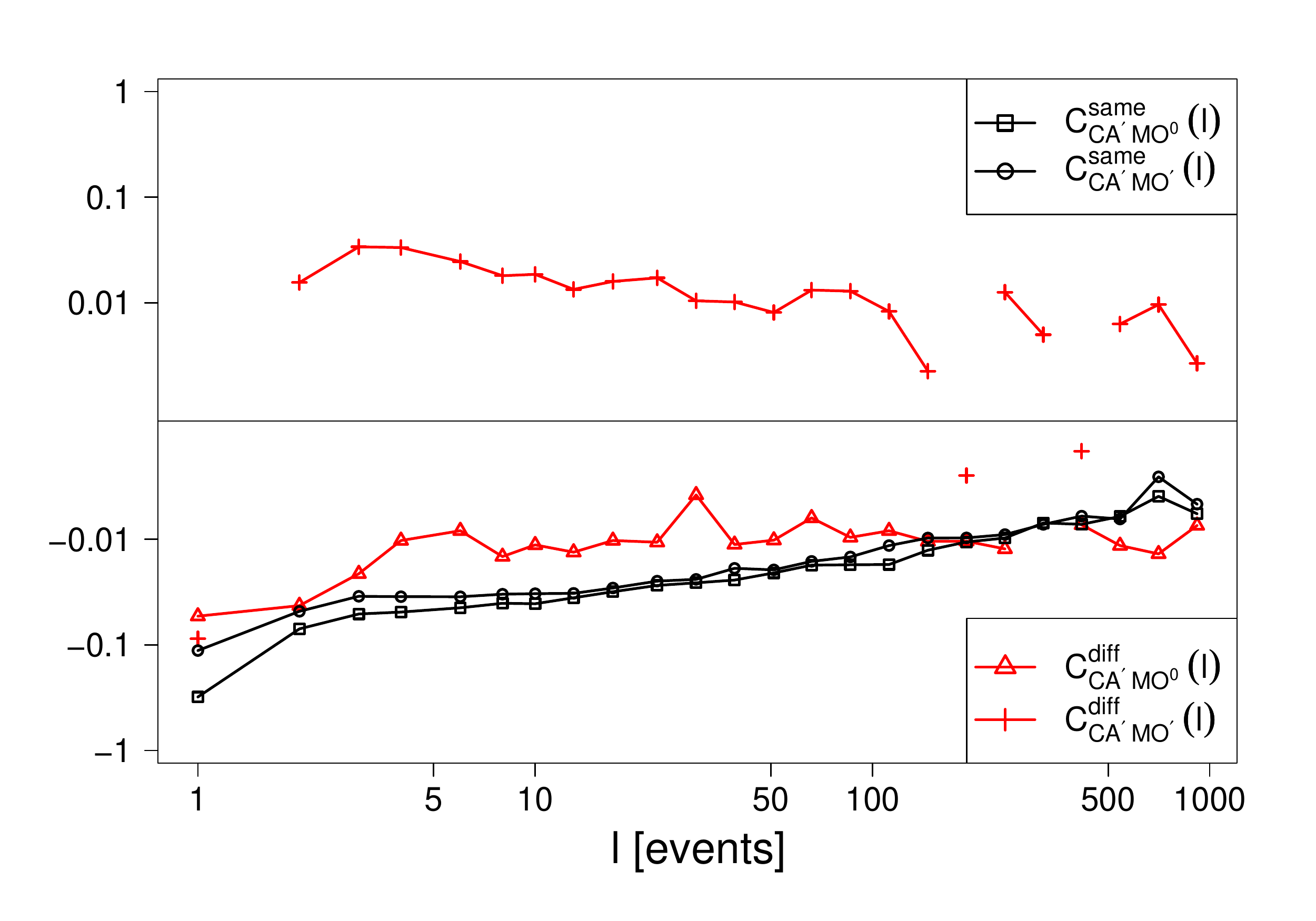}
\caption{Plots of the correlation $C_{\pi_1,\MO}(\ell)$. {\it (left)} $\pi_1=\CAO$; {\it (right)} $\pi_1=\CA'$}
\label{fig7}
\end{center}
\end{figure}

\begin{figure}[h]
\begin{center}
\includegraphics[width=0.49\textwidth]{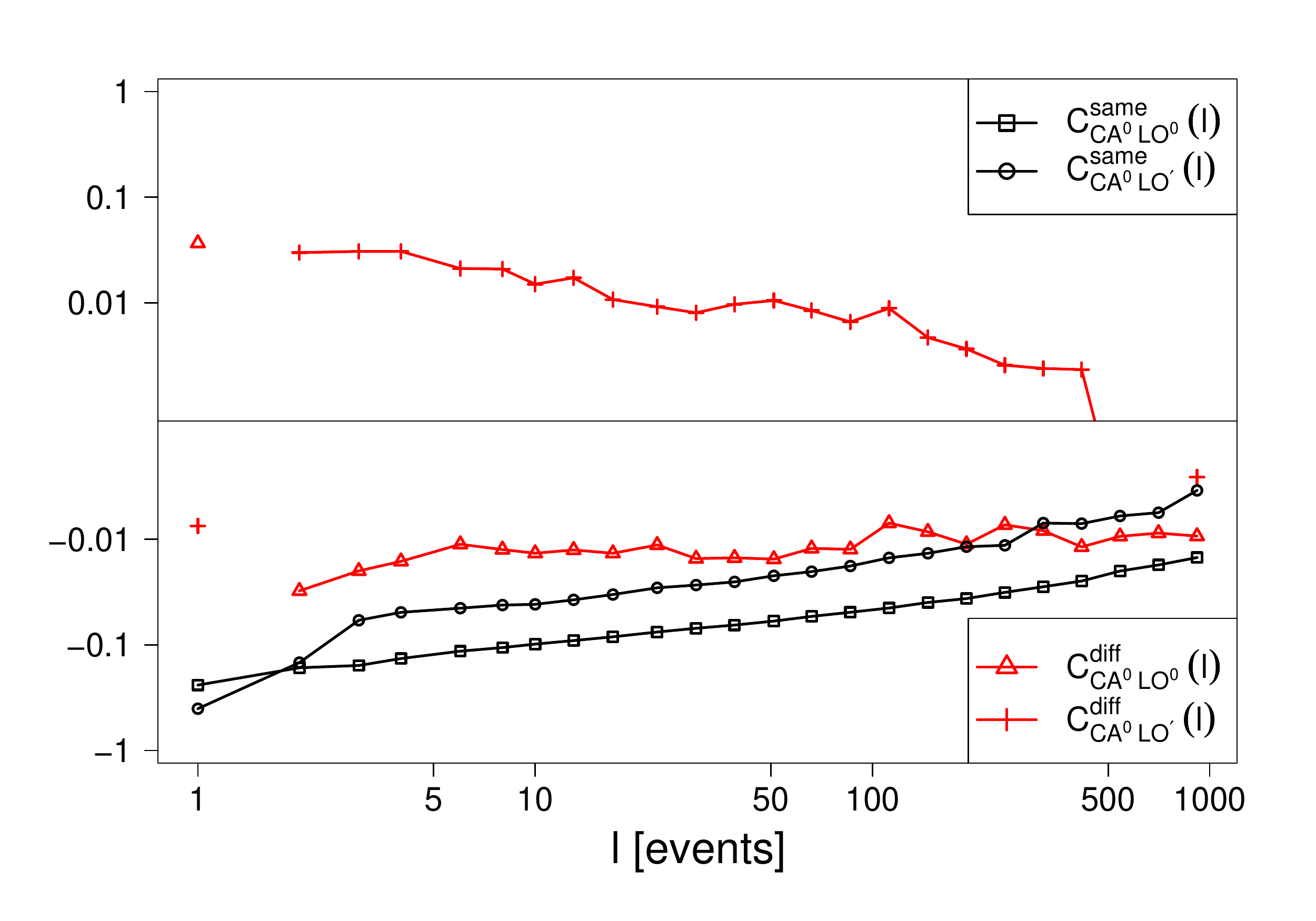}
\includegraphics[width=0.49\textwidth]{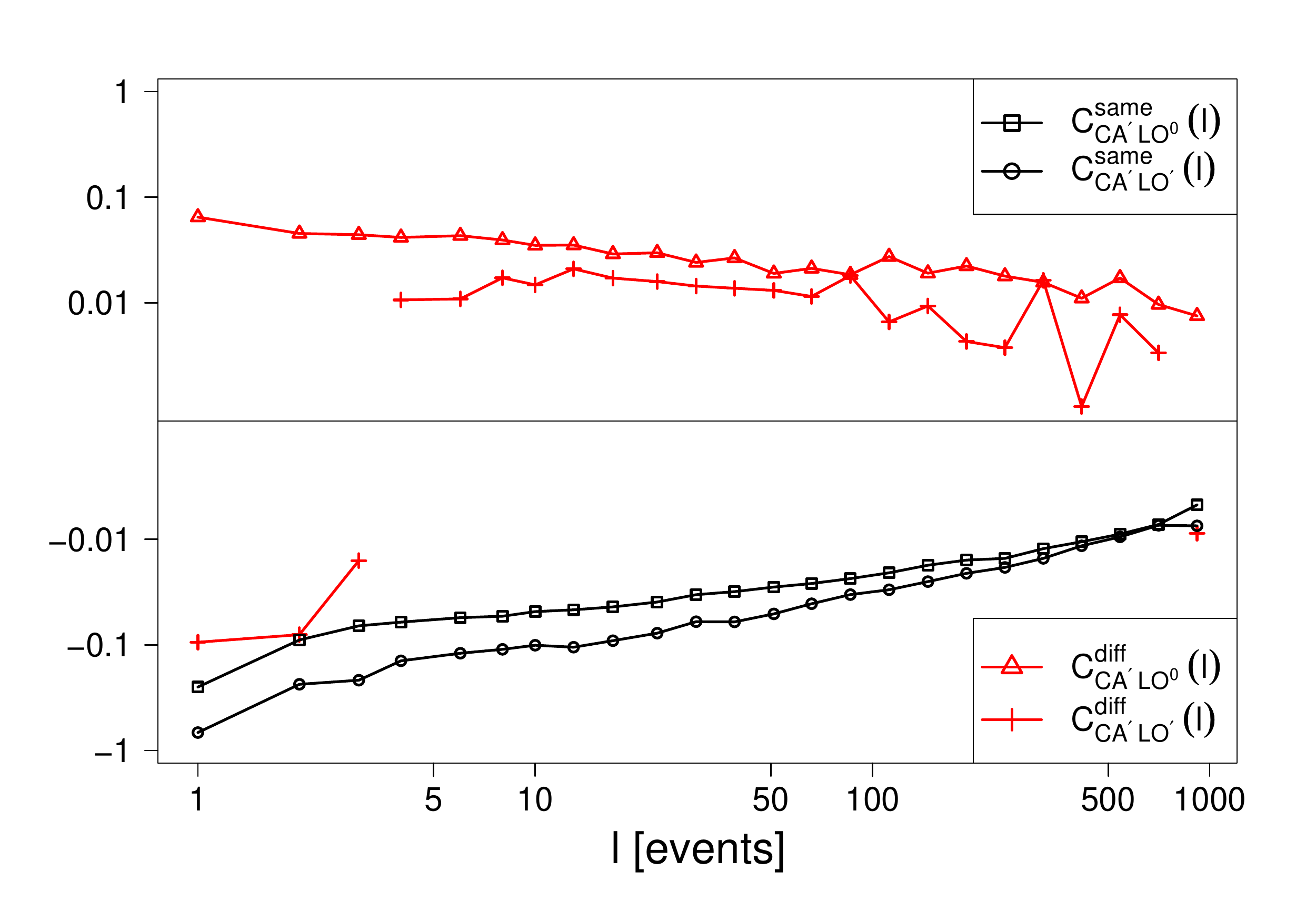}
\caption{Plots of the correlation $C_{\pi_1,\LO}(\ell)$. {\it (left)} $\pi_1=\CAO$; {\it (right)} $\pi_1=\CA'$}
\label{fig8}
\end{center}
\end{figure}

\begin{figure}[h]
\begin{center}
\includegraphics[width=0.49\textwidth]{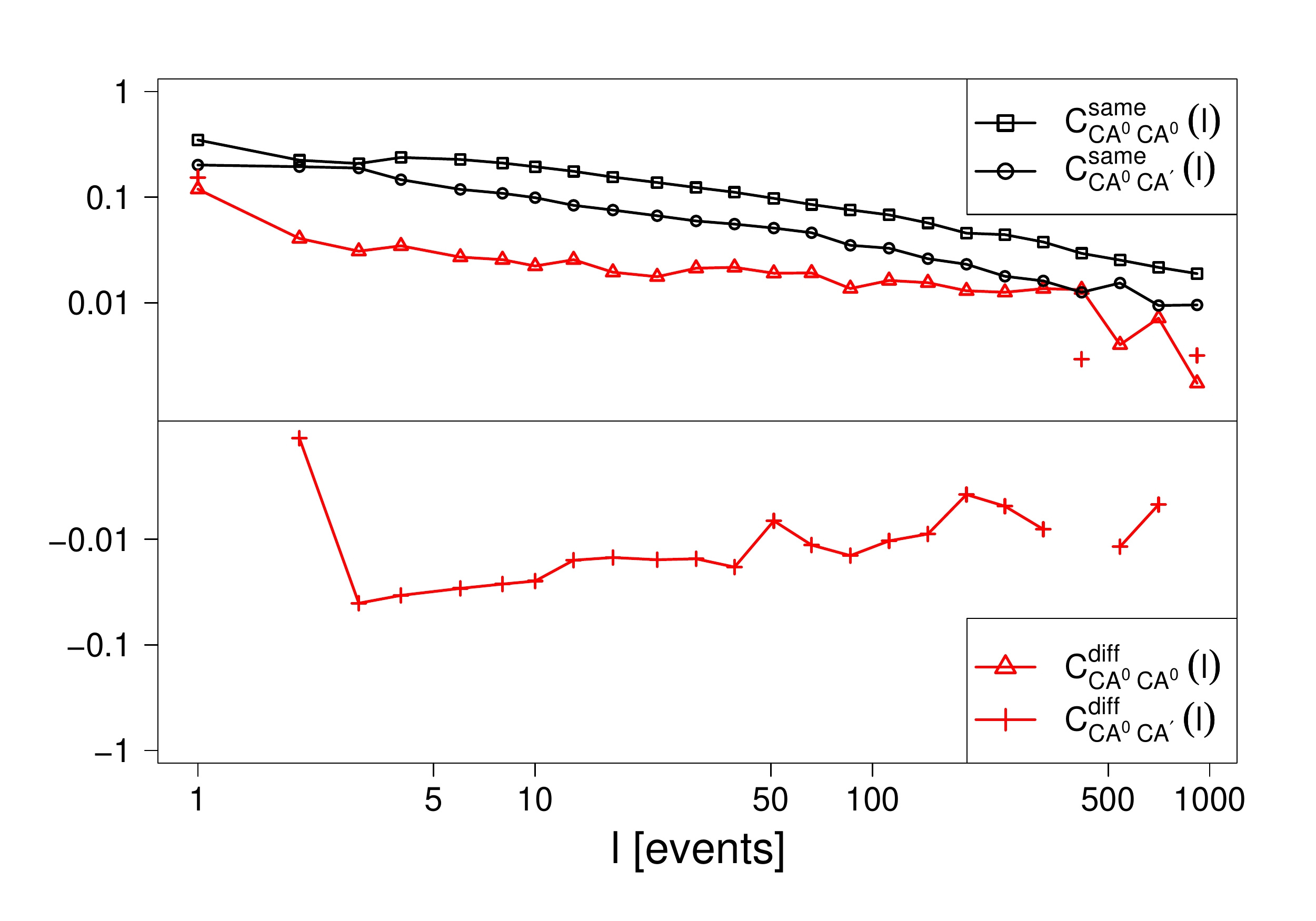}
\includegraphics[width=0.49\textwidth]{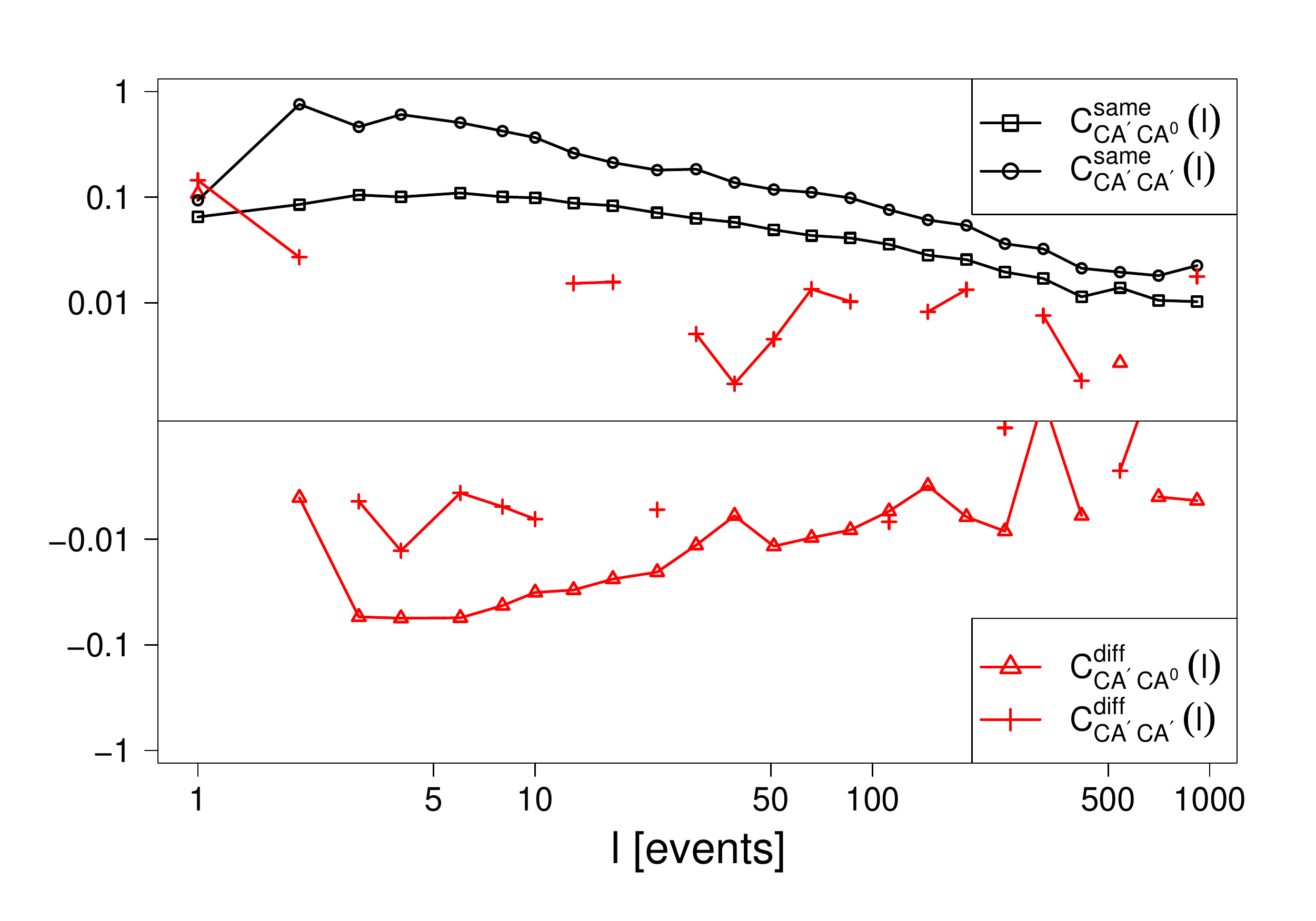}
\caption{Plots of the correlation $C_{\pi_1,\CA}(\ell)$. {\it (left)} $\pi_1=\CAO$; {\it (right)} $\pi_1=\CA'$}
\label{fig9}
\end{center}
\end{figure}

\end{document}